\documentclass[review]{elsarticle}
\usepackage{natbib}
\usepackage{graphicx}
\usepackage[hidelinks]{hyperref}
\usepackage{here}
\usepackage{multicol}
\usepackage{multirow}
\usepackage{amsmath, amssymb}
\usepackage{amsthm}
\usepackage{calc}
\usepackage{color}
\usepackage[dvipsnames]{xcolor}
\usepackage{url}
\usepackage{booktabs}
\usepackage{tabularx}  

\newcommand{\revision}[1]{\textcolor{black}{#1}}
\biboptions{authoryear}
\usepackage{tikz}

\tikzset{
  treenode/.style = {shape=rectangle, rounded corners,
                     draw, align=center,
                     top color=white, font=\small},
  root/.style     = {treenode, font=\small},
  env/.style      = {treenode, font=\small},
  dummy/.style    = {circle,draw}
}
\makeatletter
\def\ps@pprintTitle{%
 \let\@oddhead\@empty
 \let\@evenhead\@empty
 \def\@oddfoot{\centerline{\thepage}}%
 \let\@evenfoot\@oddfoot}
\makeatother

\begin{document}

\begin{frontmatter}

\title{Combining Anatomical and Functional Networks \\ for Neuropathology Identification: \\ A Case Study on Autism Spectrum Disorder}

\author{Sarah Itani\fnref{label1,label2}\corref{cor}}
\ead{sarah.itani@umons.ac.be}

\author{Dorina Thanou\fnref{label3}}
\ead{dorina.thanou@epfl.ch}

\cortext[cor]{\raggedright Corresponding author. University of Mons, Department of Mathematics and Operations Research, Rue de Houdain, 9, 7000 Mons, Belgium.}
\address[label1]{Fund for Scientific Research - FNRS (F.R.S.-FNRS), Brussels, Belgium} 
\address[label2]{Department of Mathematics and Operations Research, Faculty of Engineering,\\ University of Mons, Mons, Belgium} 
\address[label3]{Swiss Data Science Center, EPFL and ETH Z\"{u}rich, Switzerland}

\begin{abstract}
While the prevalence of Autism Spectrum Disorder (ASD) is increasing, research \revision{continues in an effort to identify common etiological and pathophysiological bases}. In this regard, modern machine learning and network science pave the way for a better understanding of the neuropathology and the development of diagnosis aid systems. The present work addresses the classification of neurotypical and ASD subjects by combining knowledge about both the \revision{structure} and the functional activity of the brain. In particular, we model the brain structure as a graph, and the resting-state functional MRI (rs-fMRI) signals as values that live on the nodes of that graph. We then borrow tools from the emerging field of Graph Signal Processing (GSP) to build features related to the frequency content of these signals. In order to make these features highly discriminative, we apply an extension of the Fukunaga-Koontz transform. Finally, we use these new markers to train a decision tree, an interpretable classification scheme, which results in a final diagnosis aid model.  Interestingly, the resulting decision tree outperforms state-of-the-art methods \revision{on the publicly available Autism Brain Imaging Data Exchange (ABIDE) \revision{collection}}. Moreover, the analysis of the predictive markers reveals the influence of the frontal and temporal lobes in the diagnosis of the disorder, which is in line with previous findings in the literature of neuroscience. Our results indicate that exploiting jointly structural and functional information of the brain can reveal important information about the complexity of the neuropathology. 
\end{abstract}

\begin{keyword}
graph signal processing; fMRI; autism spectrum disorder; \revision{explainable artificial intelligence}. 
\end{keyword}
\end{frontmatter}

\section{Introduction}

Understanding the human brain in all its complexity has always been a great challenge. Despite the undeniable progress made in the domain, some neuropathologies, such as Autism Spectrum Disorder (ASD), are characterized by the absence of a commonly defined etiology \citep{Maximo2014}. People with ASD manifest recurring behavioral patterns; they present impairments in language and communication which impede their social interactions \citep{AutismSpeaks}. With the growing prevalence of ASD, especially in children, it becomes urgent to determine the neurophysiological bases of the disorder, and use this knowledge for an early and objective diagnosis. Towards that direction, data-driven techniques are expected to shed some light on explainable and interpretable markers that capture the complexity of the neuropathology and reveal interconnected patterns on brain activation that are related to the causes of the disorder~\citep{Vargason2020, Kassraian2016}. 

While there is no consensus on the pathophysiology of ASD, it is commonly accepted by the research community that the disorder can be partially explained by studying the brain network complex structure~\citep{Uddin2013}. \revision{This structure is usually studied from a functional or a structural connectivity perspective. Functional Connectivity (FC) is typically derived from the correlation of signals such as the Blood-Oxygen-Level-Depedent (BOLD) ones, measured by resting-state functional Magnetic Resonance Imaging (rs-fMRI). Structural Connectivity (SC) is related to the detection of white matter pathways, tracked, for example, by Diffusion Tensor Imaging (DTI). Both FC and SC can be analyzed in terms of network to reveal organizing principles of the brain that are prevalent in ASD patients~\citep{Kazeminejad2019, Tolan2018, Abraham2017, Chen20162, Rane2015, Kana2014, Maximo2014, Uddin2013, Vissers2012}. }

\revision{Given the complex nature of the brain and the paramount importance of both functional and structural connectivity, in this work, we combine both sources of information by borrowing tools from the emerging field of Graph Signal Processing (GSP)~\citep{Ortega2018, Shuman2013}.} GSP addresses the challenging problem of analyzing data living on an irregular domain,  that can be naturally represented by a graph. The data lying on the nodes of the graph are considered as signals (e.g., fMRI time-series) that have a strong dependency on the graph topology (e.g., brain structural connectivity). Indeed, GSP allows to integrate both structural and functional brain data by studying the interplay between graphs and  signals on graphs. 

\revision{In particular, our goal is to classify ASD from neurotypical (NT) subjects. The graph in our application consists of a set of brain regions of interest, i.e., graph nodes, that are connected based on the anatomical distance distance between them. This graph structure remains constant over time. In contrast, the data observed on top of the graph (i.e., graph signals) are of time-varying nature, since they are generated by the BOLD fluctuations. The problem thus boils down to the classification of time-varying graph signals. In order to achieve that, we extend the GSP tools further, by adapting them to the particular settings. First, we study the frequency behavior of each BOLD signal in time by computing its Graph Fourier Transform (GFT). This step generates a set of  time-varying GFT coefficients. These coefficients} are then merged into a single connectivity matrix from which we \revision{extract} discriminative graph frequency patterns, by using an extension of the Fukunaga-Koontz Transform (FKT). To eventually separate ASD from NT subjects, we use these features to train a decision tree,  which results in a final diagnosis aid model. \revision{Experimental results on the publicly available ABIDE collection indicate that the proposed approach outperforms state-of-the-art methods, and confirm that exploiting jointly structural and functional information can indeed bring significant gain both in terms of performance and better understanding of the disorder.}  

\subsection{\revision{Related work}}
\revision{Structural and functional information of the brain in ASD have already been extensively studied in the literature. For the sake of completeness, we mention here some representative works in the area.
\cite{Chen20162} investigate resting-state FC in ASD over two frequency bands, namely the \textsc{Slow-4} (0.01-0.027 Hz) and \textsc{Slow-5} (0.027-0.073 Hz) bands. The FC values are used to train a Support Vector Machine (SVM) classifier. Interestingly, most of the discriminative features are located in the \textsc{Slow-4} band. Moreover, the analysis of the classification weights shows that the connections of the thalamus are the most discriminative. The work of~\cite{Goch2014} uses structural network features such as the clustering coefficient and the betweeness centrality to achieve the classification of typically developing and ASD subjects through SVM. The achieved accuracy suggests the relevance of these structural features to understand the pathology. These findings are confirmed by~\cite{Tolan2018}, who compute graph measures on both fMRI and DTI-based networks at both global (e.g. diameter, modularity) and node levels (e.g., eccentricity, path length). The features  are then used  to achieve successful predictions through ensemble classification models consisting of SVM, decision tree, and $K$-nearest neighbor classifiers. Finally, a few studies have considered the joint analysis of both FC and SC, in quest of common functional and structural patterns of (hypo- or hyper-) connectivity in given brain areas~\citep{Ray2014, Mueller2013, Nair2013, Uddin2013}. Such an example is the study of~\cite{Mueller2013}, which reveals common functional and structural impairments in the right temporo-parietal junction area, the left frontal lobe, and the bilateral superior temporal gyrus.}

\revision{More recently, tools from the GSP framework have been proposed to combine structural and functional information in many brain related applications. Most of them focus on  the analysis and classification of BOLD signals extracted from fMRI data, through the popular  GFT. However, using GFT coefficients in their simple form is clearly not enough to capture activation patterns of the brain.  For that reason, some studies perform the decomposition of signals into three components (i.e., low, medium and high frequency) in order to find patterns which are statistically significant to characterize cognitive flexibility~\citep{Medaglia2018} and motor skill~\citep{Huang2016}. Along the same lines,~\cite{Wang2018} suggest that GFT coefficients are not sufficient classification patterns, though used as inputs of a high-performing algorithm such as SVM. The work of~\cite{Ktena2018} shows that GFT coefficients have better discriminating power when they are embedded within more complex patterns, learned through deep learning architectures such as convolutional neural networks. These findings suggest that the discriminative information brought by the time-varying GFT coefficients might be hidden in complex patterns. Thus, the challenge remains to  discover these complex patterns, and at the same time, ensure  a certain level of \textit{interpretability}, i.e., the obtained patterns should be understood to a certain extent by humans, and more particularly clinicians in the context of medical data~\citep{Itani2019, Doshi2017}.}

\subsection{\revision{Contributions}}
The contributions of our study are summarized as follows. 
\begin{itemize}
    \item [(1)] \revision{We extend the GSP framework by proposing a way of discriminating time-series in the graph Fourier domain. In that respect, we show that ASD can be predicted based on frequency patterns on the structural graph that are sophisticated, while remaining interpretable. Our results suggest  that the differences between the ASD and NT subjects cannot be clearly attributed to specific graph frequency bands (i.e., typically low, middle and high bands), contrarily to what was previously put forward \citep{Medaglia2018, Menoret2017, Huang2016}.} 
    
    \item [(2)]\revision{Our discriminative patterns can be classified by a decision tree without the necessity of using more complex classification schemes, e.g., through deep learning as was considered by~\cite{Ktena2018}. This addresses to some extent the challenge of interpretability which arises for the development of diagnosis aid models~\citep{Itani2019, Doshi2017}.} 
    
    \item [(3)] In terms of classification accuracy, our framework outperforms other state-of-the-art methods that are based on either  the structural or the functional connectivity, i.e.,  the Graph Fourier Transform - GFT~\citep{Ortega2018} and the Spatial Filtering Method - SFM~\citep{Subbaraju2017}. Moreover, the interpretation of the results confirms previous findings of the neuroscience literature about ASD.  
\end{itemize}

\revision{
The remainder of the paper is structured as follows. First, we present the materials and methods for our study in Sec.~\ref{sec:stateOfTheArt}. Then, we describe our experimental protocol in Sec.~\ref{sec:exProto}. We expose our results in Sec.~\ref{sec:R&D}, which are followed by a discussion in Sec.~\ref{Interpretation}. Finally, we conclude the paper in Sec.~\ref{sec:conclusion}. }

\section{Materials and methods}~\label{sec:stateOfTheArt}
\begin{figure}
    \centering
    \includegraphics[scale=0.45]{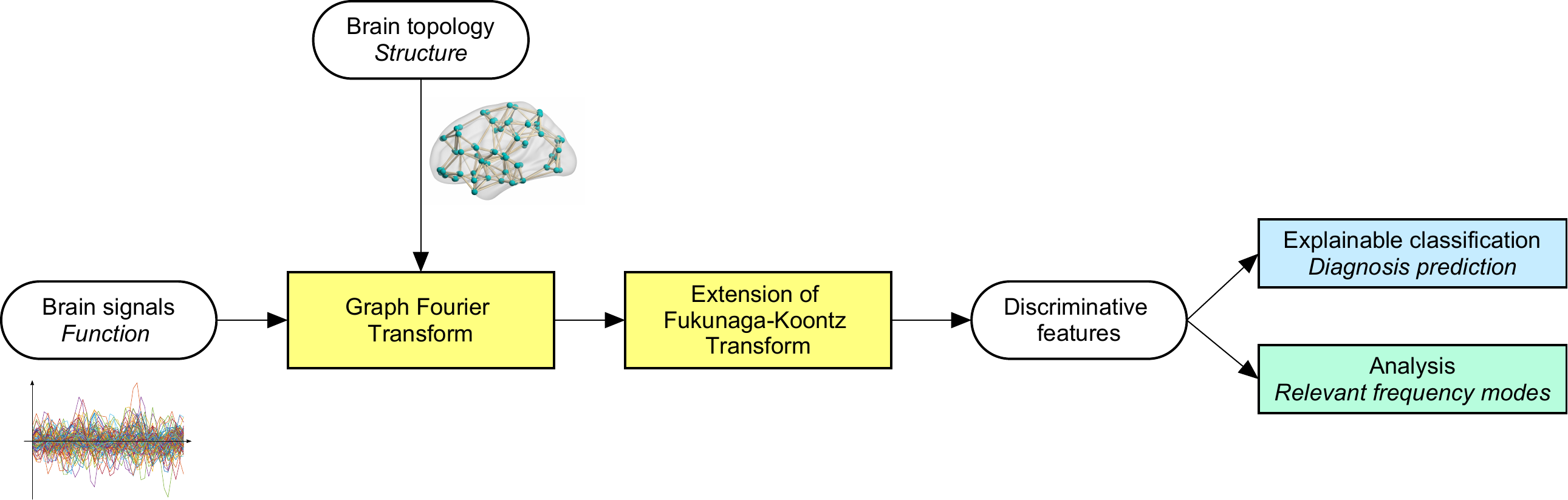}
    \caption{Overview of the proposed approach.}
    \label{GraphicalAb}
\end{figure} 
\revision{In this section, we present our novel graph-based framework for the classification of the BOLD time-series (see Fig.~\ref{GraphicalAb}), by describing in detail the tools and the building blocks.} \revision{For the sake of simplicity, first, we present the notation that is used in the remainder of the paper (Subsec.~\ref{sec:notations}). Then, in Subsec.~\ref{sec:GFT},  we introduce a methodology for jointly modeling the time-varying BOLD data and the underlying brain structure, by extending existing tools from GSP to our settings. Furthermore, in order to discriminate the BOLD time-series, in Subsec.~\ref{sec:ourFramework}, we extend the Fukunaga-Koontz transform  to the graph-based representation of the previous subsection. \revision{Finally, we present the methodology used for classification and analysis (Subsec.~\ref{sec:lastStep}).}}

\subsection{Settings and notations}\label{sec:notations}

The classification problem that we tackle in this work consists of two classes: (i) NT, and (ii) ASD subjects. We denote by:
\begin{itemize}
\item $n_T$, the total number of subjects; 
\item $n_A$, the total number of ASD subjects; 
\item $n_N$, the total number of NT subjects. 
\end{itemize}
Blood-Oxygen-Level-Dependent (BOLD) signals are available for each subject\revision{; they are provided for a set of $r$ Regions Of Interest (ROIs)}. For each subject $i$ ($i= \{1,\ldots n_T$\}), we denote by: 
\begin{itemize}
\item $T_i$, the number of time-points in the BOLD time-series;
\item $\mathbf{X}_i$, the $r\times T_i$ matrix of BOLD time-series.  
\end{itemize}

\subsection{\revision{Graph-based representation of BOLD time-series}}\label{sec:GFT}
We model the structure of the brain as an undirected, weighted graph $\mathcal{G}$, where the set of nodes $\nu$ correspond to the brain regions of interest. The edges of the graph are defined by connecting close ROIs in terms of their topological distance in the brain. In particular, we define the weight $A_{uv}$ between two nodes $u, v$ of a brain graph $\mathcal{G}$ as the inverse of the distance $d_{uv}$ between the two nodes (i.e., regions of interest). Thus, the adjacency matrix $\mathbf{A}$ is such that:  
\begin{align*}
\label{eq:weights} A_{uv} = d_{uv}^{-1} \:\:\:\:\text{and}\:\:\:\: A_{uu} = 0 \:\:\:\:\text{for}\:\:\:\: u,v = 1,\ldots r.
\end{align*}
For each node, we keep only  its $K$ nearest neighbors, while ensuring that the final graph is symmetric. The final adjacency matrix is computed as: 
\begin{equation}\label{eq:adjacencyMat}
\mathbf{A'} = \frac{\mathbf{A}+\mathbf{A}^{T}}{2}. 
\end{equation}
For the sake of simplicity, in what follows, we denote this matrix as $\mathbf{A}$. Note that the nearest-neighbor strategy is a good proxy for representing the brain topology~\citep{Alexander2012, Bullmore2009}. \revision{The combinatorial Laplacian operator~\citep{Shuman2013} is defined as $\mathbf{L} = \mathbf{D} - \mathbf{A}$ where $\mathbf{A}$ is the graph adjacency matrix, and $\mathbf{D}$ is a diagonal matrix containing the degree of each node, i.e., $\mathbf{D}_{kk} = \sum_{j} \mathbf{A}_{kj}$.} . 

\begin{figure*}[t]
\centering
\includegraphics[scale=0.45]{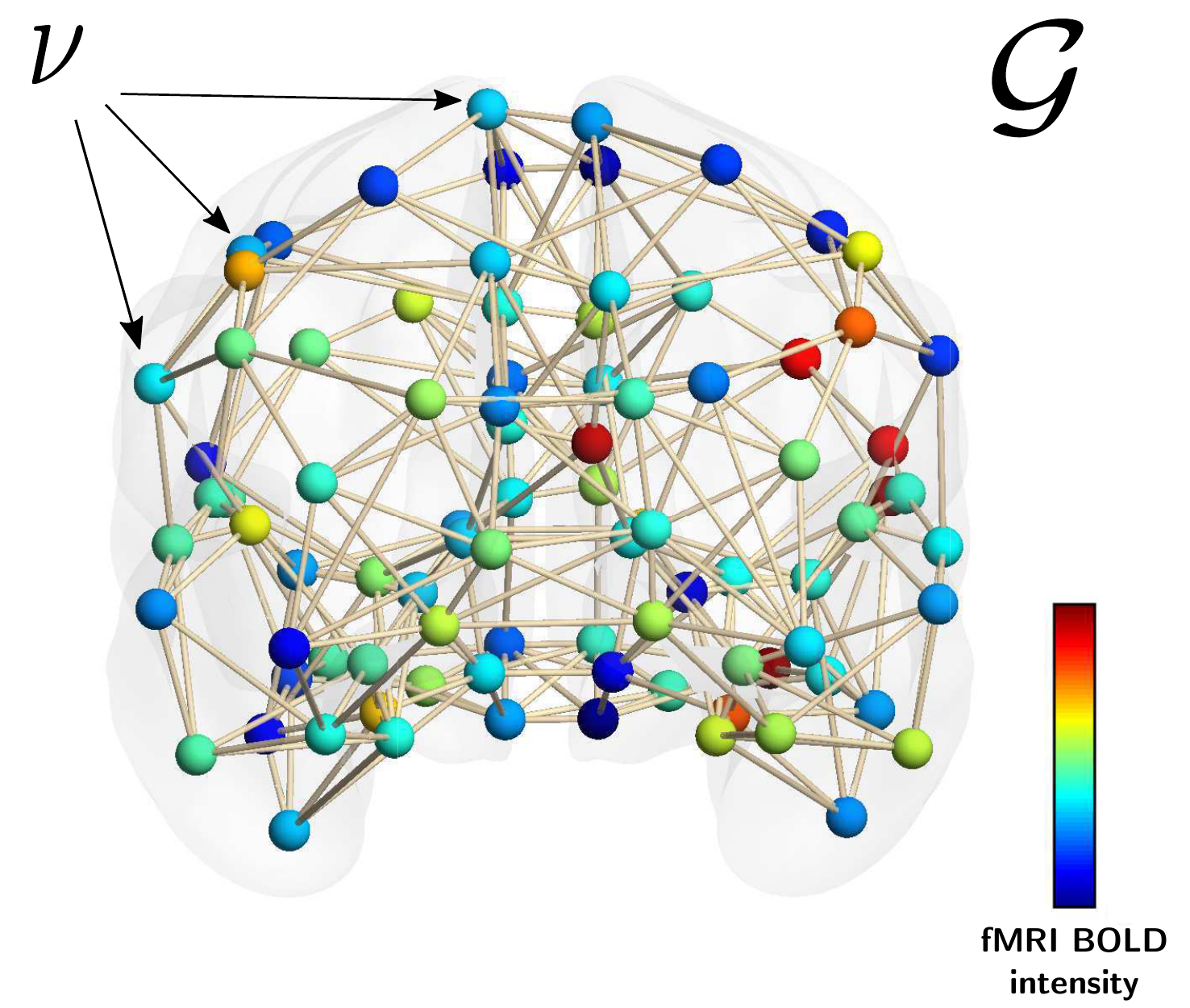} 
\caption{Representation of a graph signal, at a given time point.\protect\footnotemark}\label{fMRIsignal}
\end{figure*}
\footnotetext{Note that the brain figures presented in the present paper were drawn with the \textsc{BrainNet Software}~\citep{Xia2013}.}

\revision{Let us consider the activity of the brain at a given time instance. This involves observing the BOLD time-series issued by the ROIs at a given time point, i.e., a set of $r$ values contained in one of the column vectors of matrix $\mathbf{X}_i$ (see Subsec.~\ref{sec:notations}). This set of values lying on the brain graph is usually denoted as a \textit{graph signal}. Fig.~\ref{fMRIsignal} illustrates a 5-nearest neighbor topology of the brain, on top of which a graph signal is observed as an instantaneous measure of the brain activity. }

\begin{figure*}[t]
\centering
\includegraphics[scale=0.2]{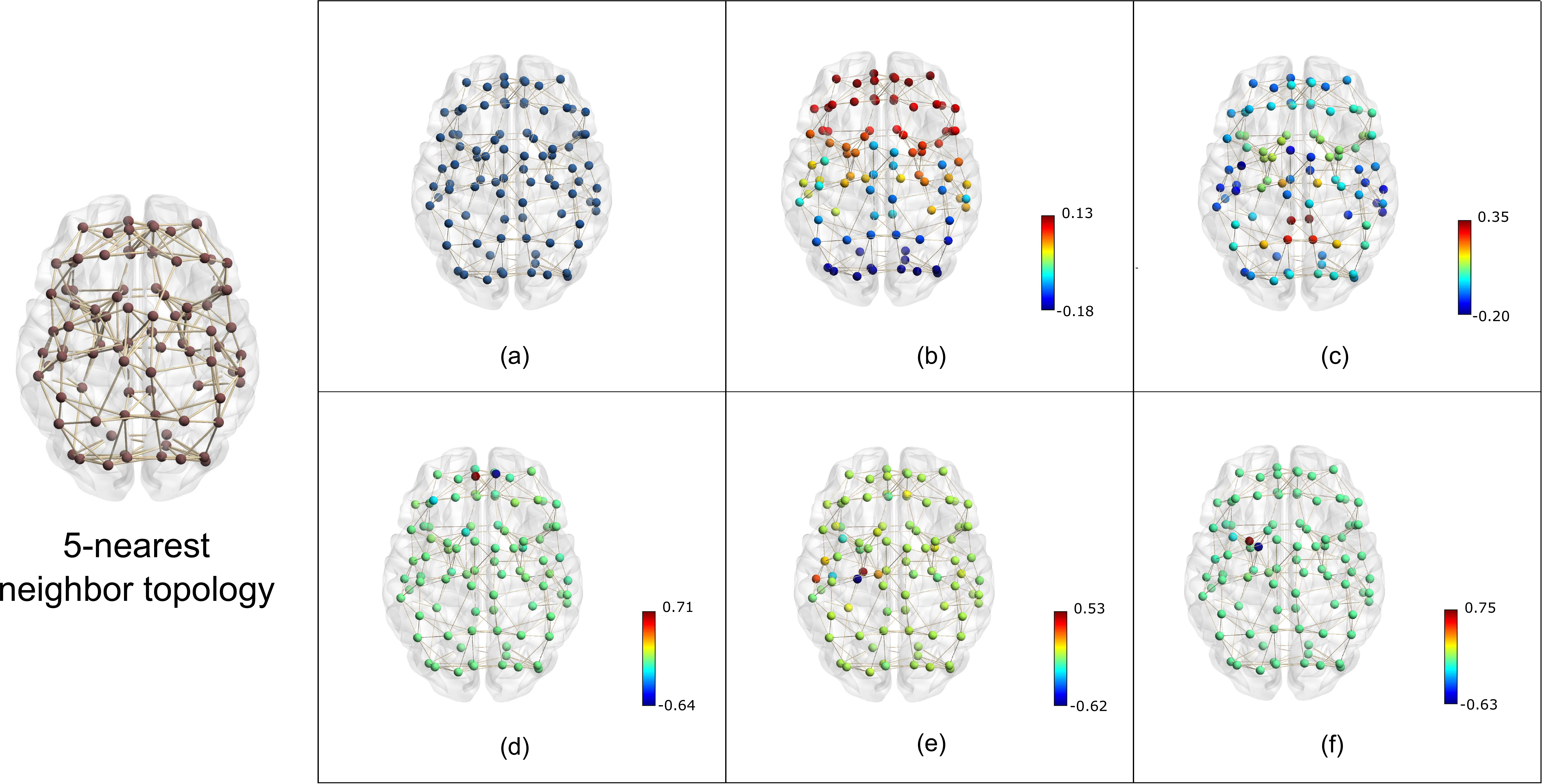} 
\caption{GFT basis illustrated on a 5-nearest neighbor topology consisting of 90 nodes~\citep{Tzourio2002}. The GFT modes (or frequencies) presented from (a) to (f) correspond to modes 1, 2, 10, 70, 75 and 90 respectively. The color represents the intensity of the signal (eigenvector of the Laplacian) on the graph. The highest the GF mode, the more fluctuating the signal values on neighboring nodes.
}\label{GFTscheme}
\end{figure*}

\revision{The spectral domain representation can reveal significant information about the characteristics of graph signals.  In particular, the Graph Fourier Transform (GFT) provides a frequency analysis of these signals, that is based on the graph Laplacian operator. The matrix of eigenvectors $\mathbf{V}$ of the Laplacian operator can be used to perform a} harmonic analysis of signals that live on the graph (i.e., graph signals), and the corresponding eigenvalues carry a notion of frequency~\citep{Shuman2013}. Indeed, the eigenvectors of the Laplacian consist of the Fourier basis that can be used for analyzing graph signals. Fig.~\ref{GFTscheme} shows the GFT modes of a 5-nearest neighbor brain graph. From $(a)$ to $(f)$, we present different graph frequency modes, in ascending order of eigenvalues, i.e., graph frequencies. The eigenvectors associated with low eigenvalues (i.e., low frequency) are smooth on the graph, which implies that  they are changing slowly across nodes that are connected by an edge. Mode $(a)$ corresponds to a constant graph signal. 

\revision{In the present work, we model our time-varying graph signals through their GFT coefficients. In particular, for each subject $i$, we compute}
\begin{equation}\label{GFTcoef}
    \mathbf{\hat{X}}_i = \mathbf{V^{T}} \mathbf{X}_i.
\end{equation}
\revision{Each column of $\mathbf{\hat{X}}_i$ relates to the GFT coefficients computed at each time point. Projecting the signals in the graph Fourier domain can be considered as a way of decorrelating the signals in that basis. In order to understand the variation of each GFT coefficient over time, we normalize the GFT coefficients at a specific instance of time, in such a way that the relative importance of each frequency component is revealed. More specifically, we normalize the columns of $\mathbf{\hat{X}}_i$ by subtracting the mean of each column and dividing by its energy i.e., the $L2$ norm (see~\ref{appA}, Eq.~\ref{Normalization}).   The resulting matrices are denoted by $\mathbf{{Y}}_i$. We then merge the normalized GFT coefficients over time in order to reveal some frequency patterns of the time-series in the graph Fourier domain. We achieve that by computing an approximation of the sample covariance matrix, that is given by:}
\begin{equation}\label{eq:jointExp}
\mathbf{S}_i = \frac{\mathbf{Y}_i{\mathbf{Y}_i^{\mathbf{T}}}}{\mathrm{tr}(\mathbf{Y}_i{\mathbf{Y}_i^{\mathbf{T}}})}. 
\end{equation}

The mean joint expectancy matrix over the patients is computed as:
\begin{equation}\label{meanJE}
   \mathbf{\Bar{S}} = \frac{1}{n_T}\sum_{i=1}^{n_T} \frac{\mathbf{Y}_i\mathbf{Y}_i^{\mathbf{T}}}{\mathrm{tr}(\mathbf{Y}_i\mathbf{Y}_i^{\mathbf{T}})}. 
\end{equation}
\revision{We compute similarly the mean joint expectancy matrices for the ASD and NT subjects, which are denoted respectively as $\mathbf{\bar{S}^A}$ and $\mathbf{\bar{S}^N}$.} It follows that: 
\begin{equation}
\mathbf{\bar{S}} =  \frac{n_A}{n_T}\mathbf{\bar{S}^A} + \frac{n_N}{n_T}\mathbf{\bar{S}^N} = \frac{n_A}{n_T} \mathbf{\bar{S}^A} + \left(1-\frac{n_A}{n_T}\right) \mathbf{\bar{S}^N}. 
\label{Lmean1}
\end{equation}
For the remainder of the development, we denote by $\alpha_A$ and $\alpha_N$ respectively, the factors $\frac{n_A}{n_T}$ and $\frac{n_N}{n_T}$, and with $\alpha_N = 1- \alpha_A$. Eq. \ref{Lmean1} can be reformulated as: 
\begin{equation}
\mathbf{\bar{S}} =  \mathrm{\alpha_A} \mathbf{\bar{S}^A} +~\mathrm{\alpha_N} \mathbf{\bar{S}^N}.
\label{Lmean2}
\end{equation}
The mean joint expectancy matrix is thus expressed as a positive linear combination of both mean joint expectancy matrices of ASD and NT subjects. 

\subsection{\revision{Finding a discriminative subspace}}\label{sec:ourFramework}
\revision{After computing a representative matrix that captures the temporal evolution of the graph spectral components of the BOLD fluctuations for each category, we need to classify the subjects in one of the two categories. To achieve this, we propose an extension of the Fukunaga-Koontz Transform (FKT)~\citep{Fukunaga2013, Fukunaga1970} to time-varying graph signals. This method achieves separation of two classes by relying on the simultaneous diagonalization of the covariance matrices~\citep{Huo2004}. Such a diagonalization results in a set of common eigenvectors, along which the classes have complementary eigenvalues. This means that a significant dimension for a class is less relevant for the other, and vice-versa. Reduction is performed in keeping the dimensions associated with the highest eigenvalues for each class.
Our proposed extension of the method to our graph settings consists of the following three steps.}

\paragraph{}\textbf{Whitening}. \revision{First, we need to decorrelate the data by means of a whitening operator}. Let us consider the eigen-decomposition of the matrix $\mathbf{\bar{S}}$:
\begin{equation*}
\mathbf{Q^{T}} \mathbf{\bar{S}} \mathbf{Q} = \mathbf{\Lambda} \:\:\:\: \Leftrightarrow \:\:\:\: \mathbf{\bar{S}} = \mathbf{Q} \mathbf{\Lambda} \mathbf{Q^{T}}.
\end{equation*}
As $\mathbf{\bar{S}}$  is symmetric, it holds that $\mathbf{Q^{-1}} = \mathbf{Q^{T}}$. \revision{It can be shown that $\mathbf{\bar{S}}$ has a zero eigenvalue due to the column-wise normalization of the GFT coefficients (see~\ref{appB}). We consider, without loss of generality, that the eigenvalues of $\mathbf{\bar{S}}$ are sorted in ascend order. The diagonal matrix $\mathbf{\Gamma}$ is defined as:} 
\begin{equation*}
\Gamma_{11} = 1 \:\:\:\: \text{and} \:\:\:\: \Gamma_{ii}  = \Lambda_{ii}^{-1/2}, \:\: \text{with} \:\: i = 2,\ldots r.   
\end{equation*}
Thus, if we set $\mathbf{Q_{2}} = \mathbf{\Gamma^{T} Q^{T}}$, we have $\mathbf{Q_2\bar{S}{Q_2}^{T}} = \mathbf{\text{diagonal}[0, I_{r-1}]}$. The matrix 
$\mathbf{\text{diagonal}[0, I_{r-1}]}$ has a first diagonal element equal to 0 and the remaining ones equal to one. Using the above developments, Eq.~\ref{Lmean2} is rewritten as: 
\begin{align}
&\mathbf{Q_2\bar{S}Q_2^{T}} = \mathrm{\alpha_A} \mathbf{Q_2\bar{S}Q_2^{T}} + ~\mathrm{\alpha_N}  \mathbf{Q_2\bar{S}^N Q_2^{T}} \nonumber \\
&\Leftrightarrow \mathbf{\text{diagonal}[0, I_{r-1}]}=   \mathrm{\alpha_A} \mathbf{\bar{S}^{A'}}  + ~\mathrm{\alpha_N} \mathbf{\bar{S}^{N'}}. \label{diag}
\end{align}

\noindent
\paragraph{}
\textbf{Simultaneous diagonalization of the whitened matrices}. \revision{Then, we have to find a transform which keeps the overall effect of whitening, while making the variance of the classes emerge in a complementary way. This operation corresponds to the simultaneous diagonalization of the joint expectancy matrices ($\mathbf{\bar{S}}$, $\mathbf{\bar{S}^A}$, $\mathbf{\bar{S}^N}$) which are actually related by Eq.~\ref{Lmean2}. The diagonalization is based on the results of the Newcomb's theorem (see~\ref{appC}), and performed through a matrix $\mathbf{T_2}$ such that Eq.~\ref{diag} can be reformulated as:}
\begin{align*}
 \mathbf{T_2^{T}~\text{diagonal}[0, I_{r-1}]~T_2} &=   \mathrm{\alpha_A}  \mathbf{T_2^{T}  \bar{S}^{A'} T_2} + ~\mathrm{\alpha_N}  \mathbf{T_2^{T} \bar{S}^{N'} T_2}  
 \\ 
\Leftrightarrow \: \mathbf{\text{diagonal}[0, I_{r-1}]} &= \mathrm{\alpha_A} \mathbf{\bar{S}^{A^{''}}} + \mathrm{\alpha_N} \mathbf{\bar{S}^{N^{''}}.} 
\end{align*}
\revision{The matrices $\mathbf{\bar{S}^{A^{''}}}$ and $\mathbf{\bar{S}^{N^{''}}}$ are diagonal. Their non-zero eigenvalues, multiplied respectively by $\mathrm{\alpha_A}$ and $\mathrm{\alpha_N}$, are complementary and sum to unity. The details for the construction of matrix $\mathbf{T_2}$ are provided in~\ref{psdSA}.} 

\noindent
\paragraph{}
\textbf{Computation of the projection matrix}. All the above operations can be summarized through a final projection matrix $\mathbf{P}$ such that:
\begin{align*}
&\mathbf{P\bar{S}P^{T}} = \mathrm{\alpha_A} \mathbf{P\bar{S}^A P^{T}} +~\mathrm{\alpha_N} \mathbf{P \bar{S}^N P^{T}} \nonumber \\
&\Leftrightarrow \mathbf{\text{diagonal}[0, I_{r-1}]}=   \mathrm{\alpha_A} \mathbf{\bar{S}^{A''}}  +~\mathrm{\alpha_N} \mathbf{\bar{S}^{N''}} 
\end{align*}
with
\begin{equation}\label{filter}
    \mathbf{P} = \mathbf{T_2^{T}Q_2} = \mathbf{T_2^{T}\Gamma^{T}Q^{T}}. 
\end{equation}
Thus, we end up with a matrix which can be used to project each patient's matrix of time-varying GFT coefficients in a space where the ASD and NT classes have complementary mean joint expectancy matrices. This explains why the subspace is discriminative: each class may be expressed through a subset of dimensions along which the variance of the related data is high. 

\revision{We note that the above extension is inspired by the Spatial Filtering Method (SFM). The method is introduced by~\cite{Subbaraju2017} as an extension of the FKT for the classification of ASD patients, based on their rs-fMRI data. In fact, SFM addresses a two-class classification problem by projecting the BOLD time-series of the instances in a space defined by the FKT such that they are distinctively separable. We believe that adding the structural information to the SFM framework, and studying patterns in the graph Fourier domain, can provide more insight on brain behavior in ASD patients.}

\subsection{Performing classification and analysis}\label{sec:lastStep}
\revision{The projection matrix $\mathbf{P}$ obtained from the previous subsection can be used to classify the GFT coefficients of the BOLD fluctuations.  Thus, we project the normalized GFT coefficients $\mathbf{Y}_i$ (see Subsec.~\ref{sec:GFT})} into  the discriminative matrix  $\mathbf{P}$ (see Eq.~\ref{filter}):
\begin{equation}\label{FrameCoef}
\mathbf{Z}_i = \mathbf{P}\cdot \mathbf{Y}_i.
\end{equation}
Classification is then achieved by training a model on the variance of the projected GFT coefficients. \revision{Our classification scheme is based on the decision tree mainly because of the interpretability that it provides, which is crucial for diagnosis prediction.  This quality is also enhanced by the interpretability of the features. Indeed, from Eq.~\ref{FrameCoef}, we notice that the rows of the discriminative matrix act as filters on the GFT coefficients contained in each column of matrix $\mathbf{Y}_i$. The result is a weighted sum of the GFT coefficients in the projection space on each of its dimensions. Hence, the analysis of $\mathbf{P}$ reveals the graph Fourier modes that contribute the most to the discriminative features, and thus to the diagnosis predictions.} 

Finally, we note that the variance of the elements included in the first row of $\mathbf{Z}_i$ does not carry any discriminative information. Indeed, along this first dimension, both classes are associated with a zero eigenvalue (see Eq.~\ref{filter}), due to the singularity of the mean joint expectancy matrix (see Eq.~\ref{meanJE}). In other terms, circumventing this issue was achieved at the expense of a discriminative dimension in the projection space, through a transform which is not orthonormal.

\section{Experimental protocol}\label{sec:exProto}
\begin{figure}
\centering
\includegraphics[scale=0.65, angle = -90]{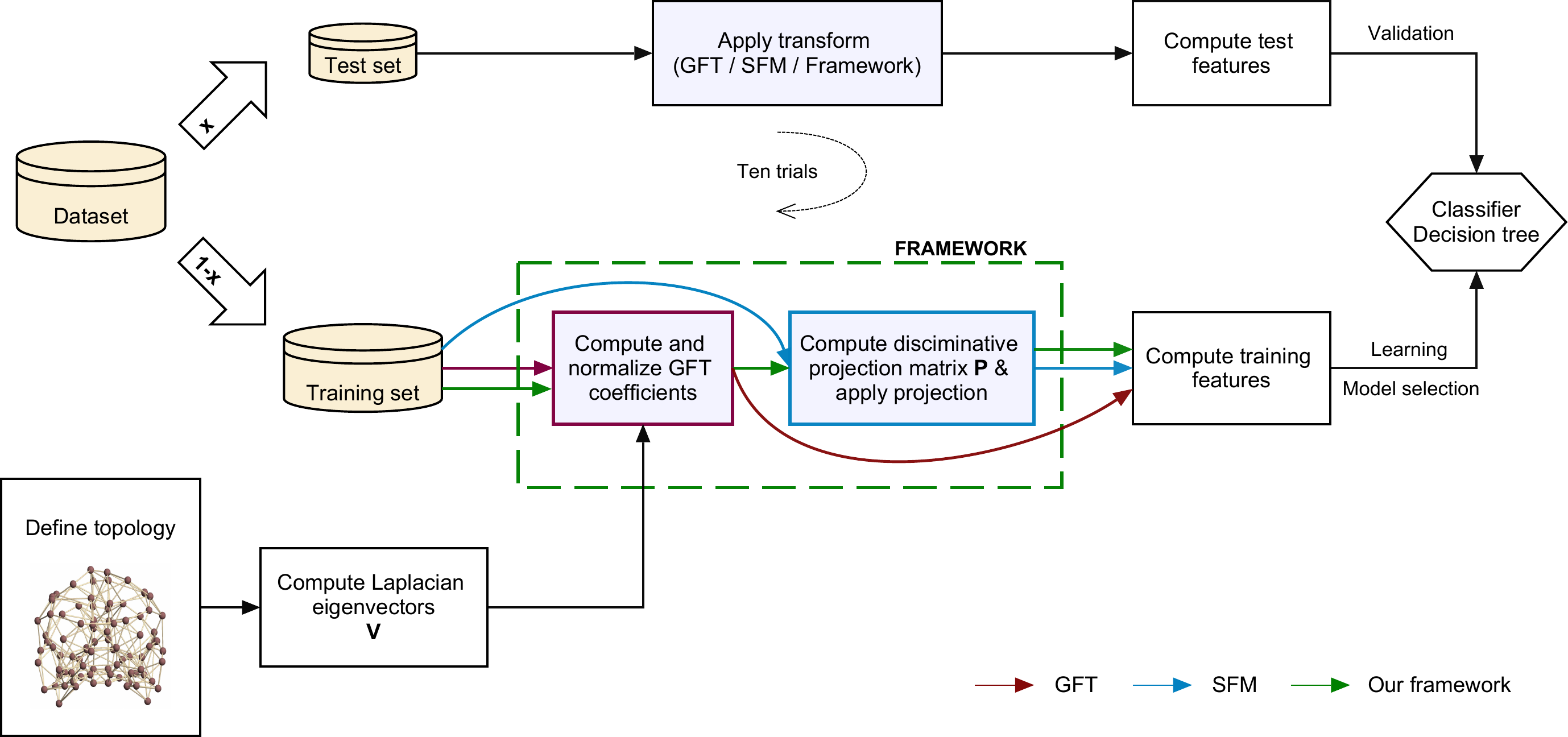}
\caption{Experimental pipeline.}\label{ExpPipe}
\end{figure}

\revision{
Fig.~\ref{ExpPipe} depicts our experimental protocol which consists of the data (Subsec.~\ref{sec3:data}), the choice of a brain topology (Subsec.~\ref{sec3:brainTop}), the assessment modalities (Subsec.~\ref{sec:AsSettings}), and a tuning strategy for classification (Subsec.~\ref{sec3:tuning}). These several aspects are introduced in the following.} 

\subsection{Data}\label{sec3:data}
In this study, we consider the ABIDE I preprocessed dataset~\citep{Di2014, Craddock2013}. It includes BOLD time-series extracted from rs-fMRI through a preprocessing pipeline which is fully detailed on the web platforms related to the ABIDE collection \citep{PCP2014, ABIDE}. The time-series considered in our work were preprocessed according to the C-PAC pipeline, which involves the following main steps: basic processing, noise signal removal, global regression, band-pass filtering (0.01-0.1 Hz), registration, time-series extraction~\citep{PCP2014}. The selected mean time-series correspond to the Automated Anatomical Labeling atlas on 90 regions of interest (AAL90), \revision{i.e., $r = 90$~\citep{Tzourio2002}. The AAL90 parcellation is documented in \ref{figInfo}}.  

Though exhaustive, the ABIDE dataset presents a high degree of heterogeneity, mainly because of the conditions under which the fMRI acquisition was operated (i.e., eyes closed/opened) as well as the demographic distribution. Thus, to ensure consistent and reliable results, we consider patients who meet the following inclusion criteria: 
\begin{itemize}
\item eyes opened during fMRI acquisition; 
\item less than 18 years old; 
\item less than 0.2 mm in mean framewise displacement. 
\end{itemize}
The corresponding data subset consists of a total of 452 subjects, with respectively 251 NT and 201 ASD subjects. 
\subsection{Brain topology definition}\label{sec3:brainTop}
\begin{figure}
    \centering
    \includegraphics[scale=0.6]{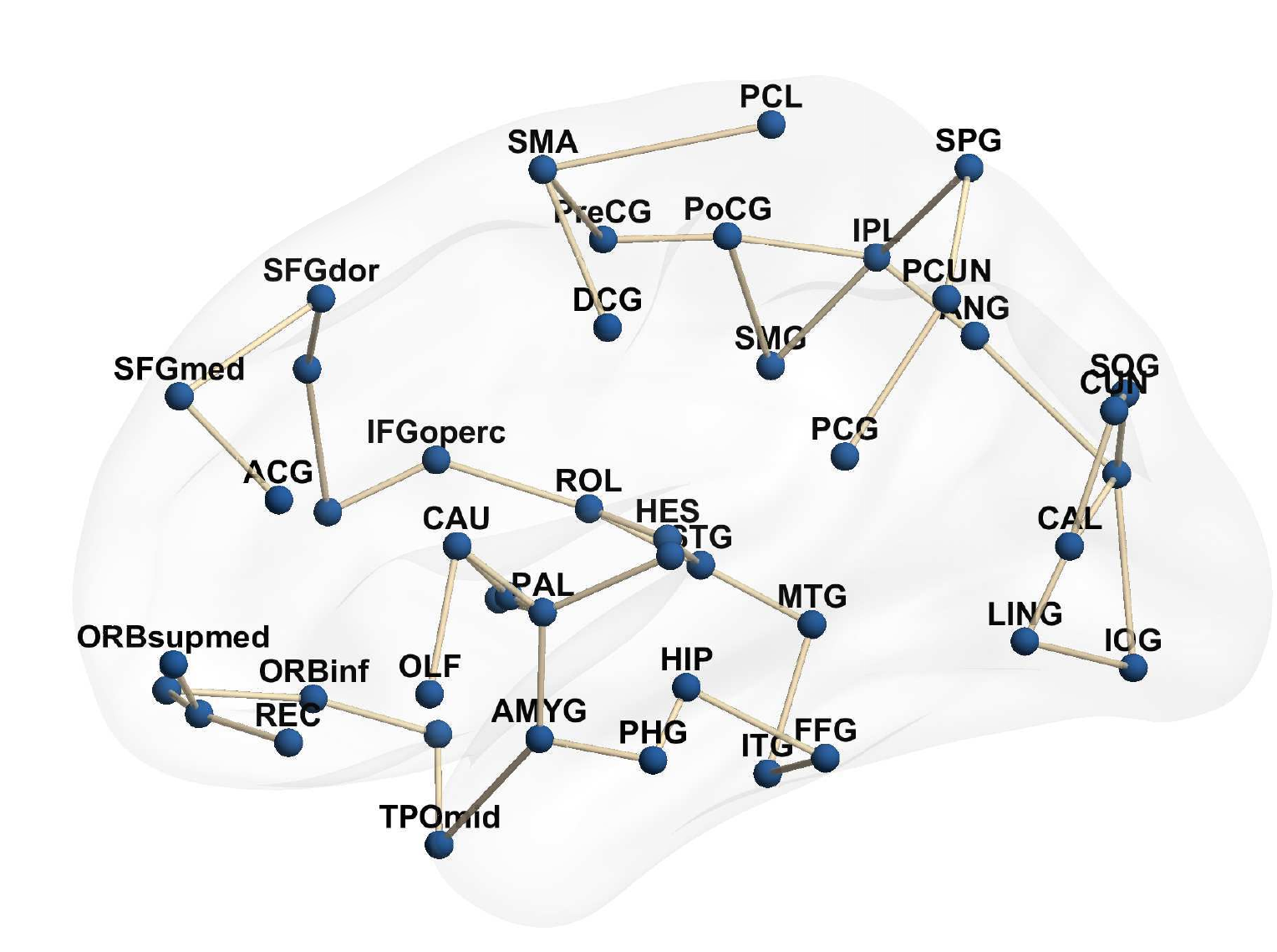}
    \caption{Two-nearest neighbor topology in the right hemisphere.}
    \label{fig:topologyK2R}
\end{figure}
As discussed in Sec.~\ref{sec:GFT}, we define the brain graph topology by following a nearest neighbor strategy. In particular, we focus on a two-nearest neighbor topology, which drew our attention in terms of spatial distribution of the ROIs. Indeed, this topology roughly divides the brain in two parts: the fronto-temporal areas on the one side, and the parieto-occipital areas on the other side (see Fig.~\ref{fig:topologyK2R}). This choice of topology is meaningful from the neuroscience point of view. Fronto-temporal areas have been associated to dysfunctions and structural abnormalities in ASD subjects~\citep{Hirata2018, Lauvin2012, Poustka2012, Goldberg1999}. Actually, the frontal lobe plays an important role in the regulation of our emotions, as it conditions our personality and our ability in making decisions~\citep{Abhang2016}. As far as the temporal lobe is concerned, it is notably involved in processing language and emotion, through the amygdala~\citep{Abhang2016, Baars2010}.

\subsection{Assessment settings}\label{sec:AsSettings}
The initial dataset is split into training and test sets (see Fig.~\ref{ExpPipe}). This is achieved by picking randomly a fraction $x$ of the total number of instances to constitute a test set. The remaining part is left for training. We consider ten trials, and report the average test accuracy. Moreover, we compare our framework with two different approaches: the first is based on the classification of the GFT coefficients (see Sec.~\ref{sec:GFT}) and the second is based on SFM. 

As shown in Fig.~\ref{ExpPipe}, within each trial, the training data are processed through GFT, SFM and our framework. For each patient, training features are derived from the projected BOLD time-series according to the following procedures. 

\begin{itemize}
    \item Concerning the GFT pipeline, we compute the variance of the normalized GFT coefficients over time. As there are 90 ROIs, there are 90 frequency modes, which result in 90 training features per time instance, per subject. 
    We also consider the set of variances related to equally-defined frequency bands, i.e., low, middle and high frequency modes. This method has been used for successful understanding of cognitive flexibility~\citep{Medaglia2018} and motor skill~\citep{Huang2016}.  
    \item For both  SFM and our framework, we consider the variance of the projected data (see Eq.~\ref{FrameCoef}) related to the $m$ most significant dimensions for each group, with $m\in [2,5]$. Thus, there is a total of $2\times m$ training features.
\end{itemize}

\subsection{Classifier tuning}\label{sec3:tuning}
In order to classify the subjects into NT and ASD we use a decision tree on the set of training features within each trial. In terms of implementation, we consider the C4.5 implementation of \textsc{Weka}~\citep{Frank2016}. All parameters are kept at their default values, except from the minimal number of instances per leaf, which is a parameter conditioning the decision tree depth. \revision{Tuning is performed through an inner cross-validation launched on each training split to select the best corresponding parameter value.}

\section{Results}\label{sec:R&D}
\subsection{\revision{Performance assessment and comparison with reference methods}}

\begin{figure}[t]
    \centering
    \includegraphics[scale=0.6]{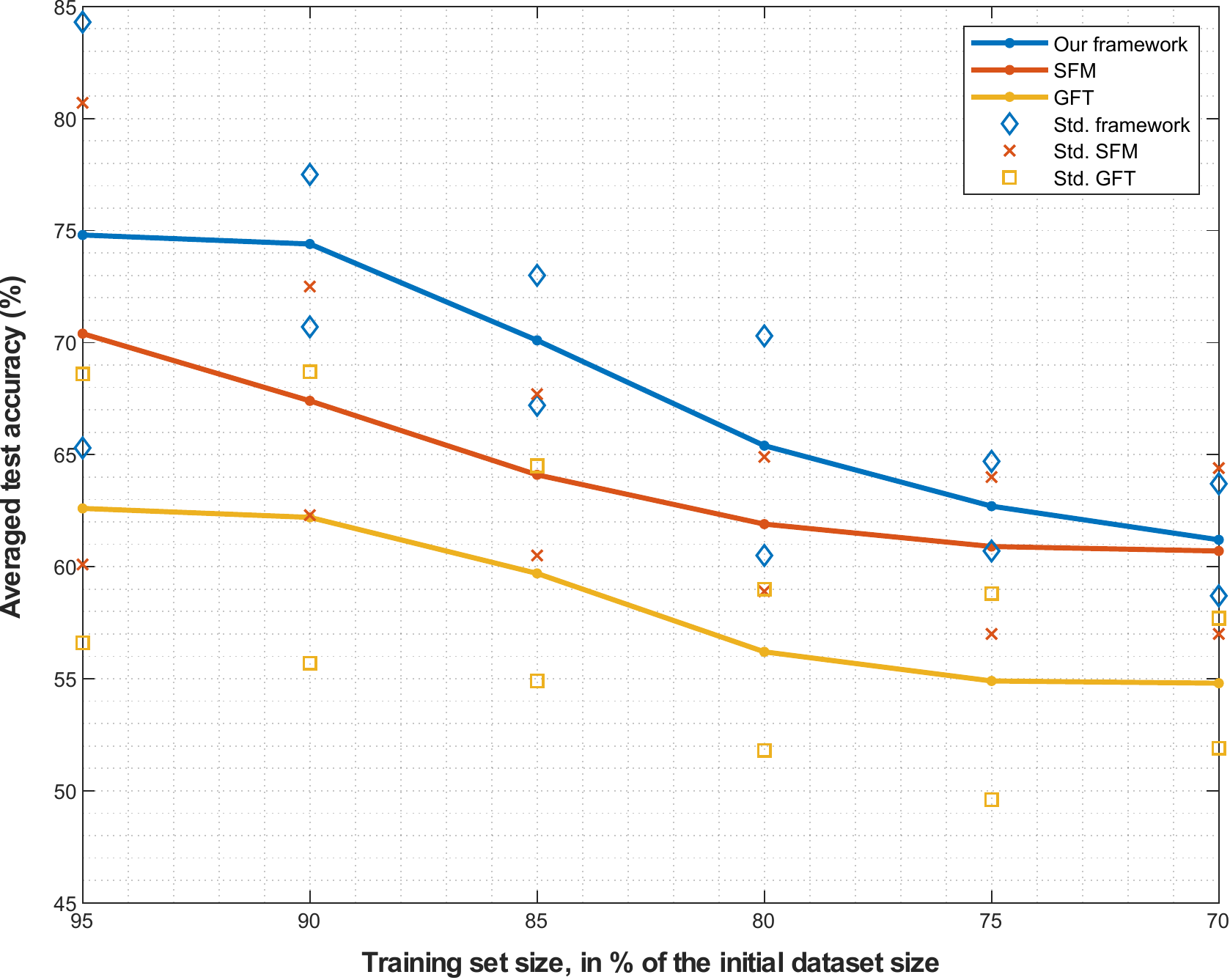}
    \caption{Performances of the methods for different training set sizes. The markers denote the standard deviation (Std.) intervals.}
    \label{fig:performances}
\end{figure}
Fig.~\ref{fig:performances} shows the best averaged test accuracies achieved by each method (across all GSP modalities), according to the procedure described in Sec.~\ref{sec:AsSettings}, for different sizes of the training set (from 95\% to 70\%, per step of 5\%). \revision{The markers around each curve delineate the standard deviation intervals related to the measures of accuracy.}  

\revision{As the training set size decreases, the performance deteriorates significantly in all the methods. Such a trend is expected, notably in the case of  SFM and our framework, which are both  based on the estimation of a covariance matrix through the sample covariance. 
Note that the standard deviation decreases as more data is available for testing.}

\revision{We draw two main observations from Fig.~\ref{fig:performances}}. 
\begin{itemize}
    \item \revision{With a performance gap of up to 12\%, our framework significantly outperforms the GFT-based approach, which suggests that the discriminative information is hidden in more complex patterns, revealed by combining the GFT coefficients. }
    \item \revision{The performance gap between SFM and our framework is less pronounced, but almost always significant. While both methods implement the FKT principle, it seems that the topological information brought by our framework influences positively the results, especially in the high-data regime.}
\end{itemize}
\revision{Table~\ref{tab:perf95} compares our approach with SFM, and presents in detail the averaged test accuracies obtained by keeping 95\% of the initial dataset for training, for different assessment modalities. We also include the $p$-values related to the Student's $t$-test. The hypothesis of equal performances is rejected for all the modalities, with $p$-values inferior to 5\%.}

\begin{table}
    \centering
    \small
    \revision{
    \begin{tabular}{l|ccc} \toprule
          \bf $m$ &  SFM (\%) & Ours (\%) & $p$-value \\ \toprule 
         2 & 69.1 $\pm$ 7.4 & \textbf{73.5} $\pm$ 6.9 & 0.026 \\ 
         3 & 70.4 $\pm$ 10.3& \textbf{74.8} $\pm$ 9.5& 0.016 \\  
         4 & 69.6 $\pm$ 10.1& \textbf{73.0} $\pm$ 6.4& $<0.001$\\ 
         5 & 67.4 $\pm$ 9.2& \textbf{71.3} $\pm$ 8.1& $<0.001$\\ \bottomrule
    \end{tabular}
   \caption{Comparison between SFM and our framework 
   in terms of averaged test accuracy.}
   \label{tab:perf95}}
\end{table}

\begin{figure}
\centering
\vspace{0.3cm}
\resizebox{\textwidth}{!}{
\begin{tikzpicture}
  [
    grow                    = down,
     level 1/.style={sibling distance=70mm},
     level 2/.style={sibling distance=40mm},
     level 3/.style={sibling distance=25mm},
     level 4/.style={sibling distance=25mm},
    level distance          = 4em,
    edge from parent path={(\tikzparentnode) -- (\tikzchildnode)}],
    every node/.style       = {font=\footnotesize},
    sloped
  ]
  \node [root,font=\scriptsize\sffamily, fill = Red, fill opacity = 0.2, text opacity=1] {ASD\_dom1}
    child { node [env,font=\scriptsize\sffamily, sibling distance = 18em, fill = ForestGreen, fill opacity = 0.2, text opacity=1] {NT\_dom2}
    	child{ node [env, font=\scriptsize\sffamily, fill = Red, fill opacity = 0.2, text opacity=1] {ASD\_dom3}
                child{ node[env, shape=circle, font=\scriptsize\sffamily, text width=0.5cm]{NT}     
    			edge from parent node [auto=right, pos=.8, font=\scriptsize\sffamily] {$\leq$ -0.14} 
    		    } 
                child{ node [env,font=\scriptsize\sffamily, sibling distance = 18em, fill = ForestGreen, fill opacity = 0.2, text opacity=1] {NT\_dom3} 
                    child{ node[env, shape=circle, font=\scriptsize\sffamily, text width=0.5cm]{ASD}     
        			edge from parent node [auto=right, pos=.8, font=\scriptsize\sffamily] {$\leq$ 0.08} 
        		    }        
                    child{ node[env, shape=circle, font=\scriptsize\sffamily, text width=0.5cm]{NT}     
        			edge from parent node [auto=left, pos=.8, font=\scriptsize\sffamily] {$>$ 0.08} 
        		    } 
    			edge from parent node [auto=mid, pos=.7, font=\scriptsize\sffamily] {]-0.14 ; 0.10]} 
    		    }     		    
                child{ node[env, shape=circle, font=\scriptsize\sffamily, text width=0.5cm]{ASD}     
    			edge from parent node [auto=left, pos=.8, font=\scriptsize\sffamily] {$>$ 0.10} 
    		    } 
    			edge from parent node [auto=right, pos=.8, font=\scriptsize\sffamily] {$\leq$ 0.06} 
    		 } 
        child{ node[env, shape=circle, font=\scriptsize\sffamily, text width=0.5cm]{NT}     
    			edge from parent node [auto=left, pos=.8, font=\scriptsize\sffamily] {$>$ 0.06} 
    		 }
      	edge from parent node [auto=right, pos=.8, font=\scriptsize\sffamily] {$\leq$ 0.04} 
          }
        child{ node[env,font=\scriptsize\sffamily, sibling distance = 18em, fill = ForestGreen, fill opacity = 0.2, text opacity=1] {NT\_dom2}
        	child{ node[env, shape=circle, font=\scriptsize\sffamily, , text width=0.5cm]{ASD}     
    			edge from parent node [auto=right, pos=.8, font=\scriptsize\sffamily] {$\leq$ 0.01} 
    		 } 
        	child{ node [env, font=\scriptsize\sffamily, fill = ForestGreen, fill opacity = 0.2, text opacity=1] {NT\_dom3}
                child{ node[env, shape=circle, font=\scriptsize\sffamily, text width=0.5cm]{ASD}     
    			edge from parent node [auto=right, pos=.8, font=\scriptsize\sffamily] {$\leq$ -0.07} 
    		    }        
                child{ node[env, shape=circle, font=\scriptsize\sffamily, text width=0.5cm]{NT}     
    			edge from parent node [auto=left, pos=.8, font=\scriptsize\sffamily] {$>$ -0.07} 
    		    } 
    			edge from parent node [auto=left, pos=.8, font=\scriptsize\sffamily] {$>$ 0.01} 
    		 } 
			edge from parent node [auto=left, pos=.8, font=\scriptsize\sffamily] {$>$ 0.04}};
\end{tikzpicture}}
\caption{Decision tree based on the projected coefficients of our framework; the subdivisions are related to the log-variance values.}\label{fig:DecisionTree}
\end{figure}
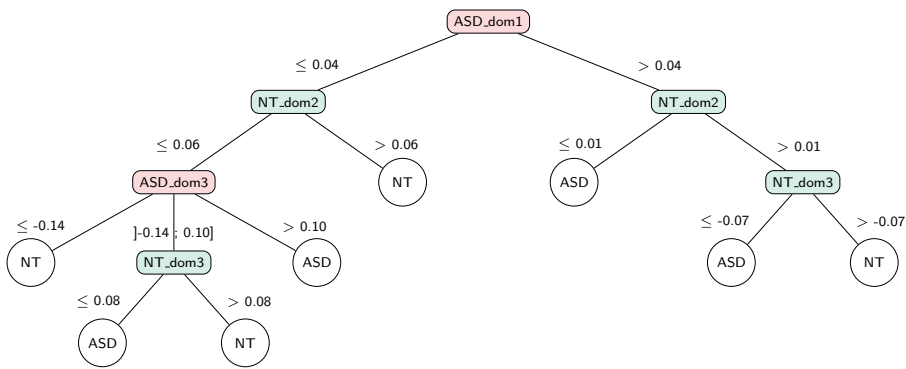

Finally, we study the interpretability of the features obtained by our framework. Fig.~\ref{fig:DecisionTree} presents the decision tree achieved on the basis of the three most discriminative dimensions ($m=3$) for each of the two groups. By \textit{ASD\_dom} (respectively \textit{NT\_dom}) we denote a dimension along which the ASD group (resp. NT group) has a large variance; the value related to this feature refers to the discriminative power of the dimension. For instance, \textit{ASD\_dom1} (resp. \textit{NT\_dom1}) refers to the first strongly dominant dimension for the ASD group (resp. NT group). The subdivisions are based on the variance (over time) of the coefficients along the dimension in question. More precisely, for a given patient, the decision tree checks the feature corresponding to \textit{ASD\_dom1}. If the variance along this dimension is high, the algorithm examines the features corresponding to \textit{NT\_dom2}. A low variance along this dimension implies an ASD diagnosis.

\subsection{Influence of the topology}
\begin{table}
    \centering
    \small
    \revision{
    \begin{tabular}{l|ccc||cc} \toprule
         \bf $m$& \bf rand WFC & \bf UC & \bf WFC & \bf SFM  & \bf 2-NN   \\ \toprule 
         \bf 2& 67.4 $\pm$ 9.4 &66.1 $\pm$ 9.5&67.8 $\pm$ 9.0& 69.1 $\pm$ 7.4 & 73.5 $\pm $ 6.9 \\ 
         \bf 3 & 66.1 $\pm$ 9.5&69.1 $\pm$ 6.3&67.8 $\pm$ 7.8& 70.4 $\pm$ 10.3& 74.8 $\pm$ 9.5\\ 
         \bf 4& 64.8 $\pm$ 7.9&65.2 $\pm$ 9.9&70.0 $\pm$ 10.4 & 69.6 $\pm$ 10.1& 73.0 $\pm$ 6.4  \\ 
         \bf 5 &67.4 $\pm$ 7.1&63.9 $\pm$ 8.3&65.2 $\pm$  11.8& 67.4 $\pm$ 9.2 & 71.3 $\pm$ 8.1  \\ \bottomrule
    \end{tabular}}
   \caption{\revision{Influence of the topology on the predictive performances}}\label{tab:perfTopology}
\end{table}

In order to understand the influence of the topology on the classification performance, we perform additional experiments, by considering alternative connectivity matrices:
\begin{itemize}
    \item \revision{a Weighted Fully Connected (\textbf{WFC}) topology which is generated by connecting all the nodes of the graph to each other, in assigning as a weight the inverse distance (see Subec.~\ref{sec:GFT}). We also consider a random WFC (\textbf{rand WFC});}
    \item \revision{a Uniformly Connected (\textbf{UC}) topology, generated by connecting all the nodes to each other, in assigning a constant and unit weight to all the connections}. 
\end{itemize}
\noindent
\revision{Table~\ref{tab:perfTopology} presents the corresponding predictive performances, which are statistically different from the results achieved with a 2-NN topology ($p <0.05$, see Table~\ref{tab:pValuesFFM}). For completeness, we show the results of the SFM method.  It appears that introducing a form of structural information leads to different results from those of the SFM method. As the accuracies reported in Table 2 suggest, the knowledge-guided choice of the 2-NN topology proves beneficial in this respect. On the contrary, the projection of the BOLD time-series over a random structure (i.e., rand WFC) or a uniformly connected structure (i.e., UC) reduces the classification performances. Finally, it is worth noting that the performances achieved with the WFC topology are inferior to those achieved in the case of 2-NN. This shows that probably the edges added to the 2-NN graph, i.e., non-local interactions, add some noise in the process. }

\begin{table}
    \centering
    \small\revision{
    \begin{tabular}{lcccc} \toprule
         \bf $m$& \bf rand WFC  & \bf UC & \bf WFC   \\ \toprule 
         \bf 2& 0.010  & 0.006 & 0.002 \\ 
         \bf 3& 0.005 & 0.002 & 0.008\\ 
         \bf 4& 0.002 & 0.003 & 0.033\\ 
         \bf 5& 0.002& 0.005 & 0.005 \\ \bottomrule
    \end{tabular}
   \caption{$p$-values related to the comparison with our 2-NN-based framework (Student's $t$-test)}\label{tab:pValuesFFM}}
\end{table}

\subsection{\revision{Results for the adult population}}
\revision{The population targeted by our case study relates to adolescents (less than 18 years old). For assessment purposes, we consider another subsample of the ABIDE dataset which includes adults. This subsample follows the same inclusion criteria as those presented in Sec.~\ref{sec3:data}, except for the age (superior to 18 years old in this case). The sample includes a total of 130 subjects, with 63 ASD and 67 neurotypical subjects. Given the limited size of the dataset, we consider a Leave-One-Out-Cross-Validation (LOOCV) procedure, and report the results for several values of parameter $m$, including those acquired using the entire training features. The results are reported in Table~\ref{tab:+18yo} for SFM and our framework based on a 2-NN topology. The methods are statistically different for all the tested modalities, with  $p$-values strictly inferior to 0.001. }

\begin{table}
    \centering
    \small
    \revision{
    \begin{tabular}{l|ccc} \toprule
          \bf $m$ &  SFM (\%) & Ours (\%) \\ \toprule 
         45 & 54.6 & \textbf{66.9} \\ 
         12 & 55.4 & \bf 67.7 \\
         9 & 52.3 & \bf 67.7 \\  
         6 & 52.3 & \bf 53.8\\ 
         \bottomrule
    \end{tabular}}
   \caption{LOOCV accuracies achieved by SFM and our framework over the adult population}\label{tab:+18yo}
\end{table}

\revision{Note that the performances achieved over the adult population are inferior to those reached on the adolescent population. This is certainly attributable to the reduced size of the adult dataset, which includes 130 subjects against 452 subjects in the adolescent sample. Yet the ASD population is quite heterogeneous in profiles given the extent of the spectrum. The appropriate representation of the neuropathology thus requires a large amount of data, especially as the adult population covers here a large range of ages (from 18 to 50 years old). Despite these unfavorable conditions, we note that our framework outperforms the SFM method, with a performance gap that reaches up to 15\%.  }
\section{Discussion}\label{Interpretation}
\revision{In this section, we discuss different aspects of the proposed framework and the achieved results. First, we analyze the discriminative features, in order to reveal and interpret the corresponding brain patterns (Subsec.~\ref{sec:interpretation}). Then, we compare our results with the state of the art for the ABIDE dataset (Subsec.~\ref{sec:compABIDE}). Finally, we discuss the limitations of the present study and propose future directions accordingly (Subsec.~\ref{sec:limitations}). }

\subsection{\revision{Interpretation of the discriminative features}}~\label{sec:interpretation}
\begin{figure}
\centering
\includegraphics[scale=0.7]{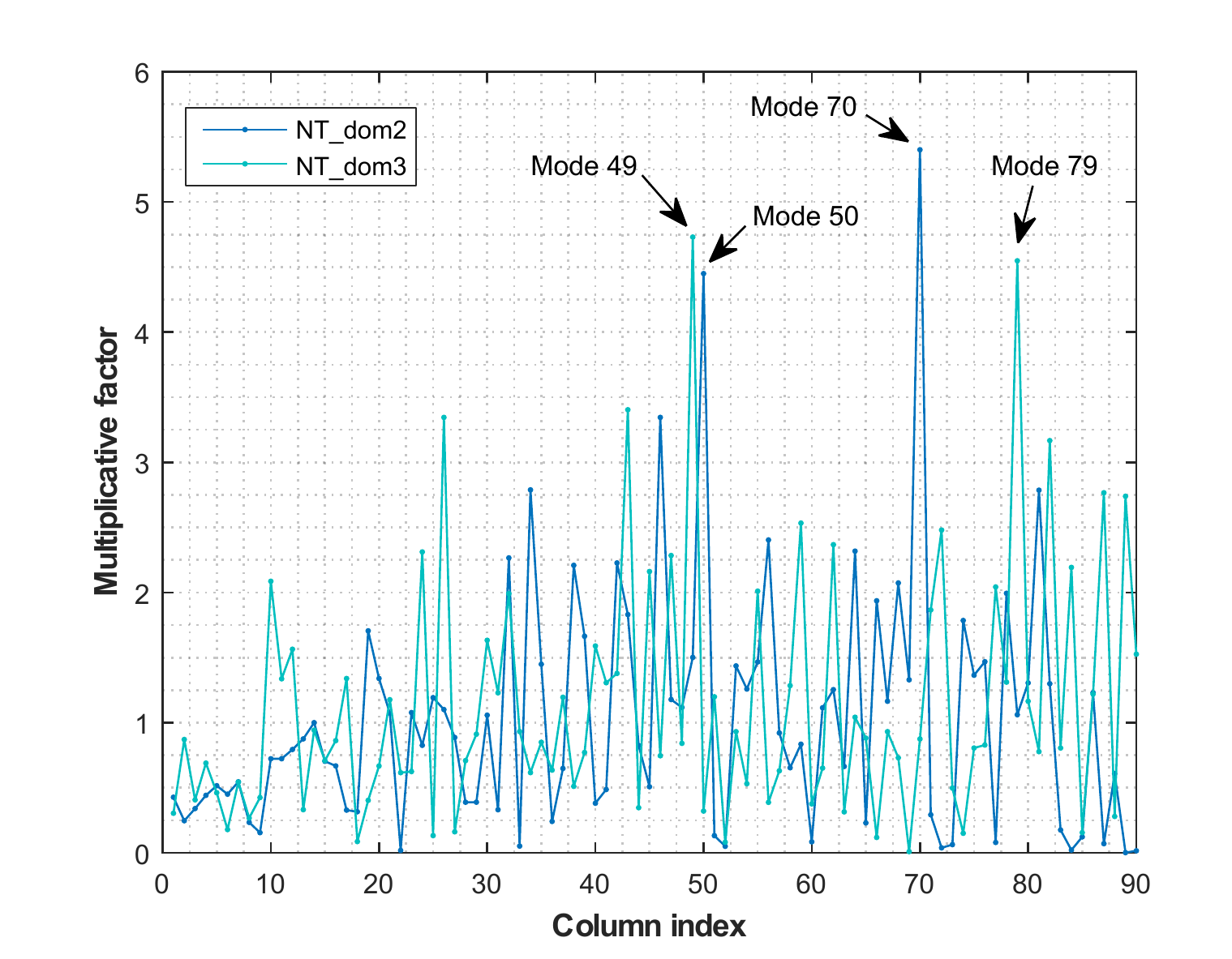}
\caption{Matrix P: interpretation of the rows -- NT-dominant dimensions.}\label{fig:matP_NTrows}
\end{figure}

\begin{figure}
\centering
\includegraphics[scale=0.7]{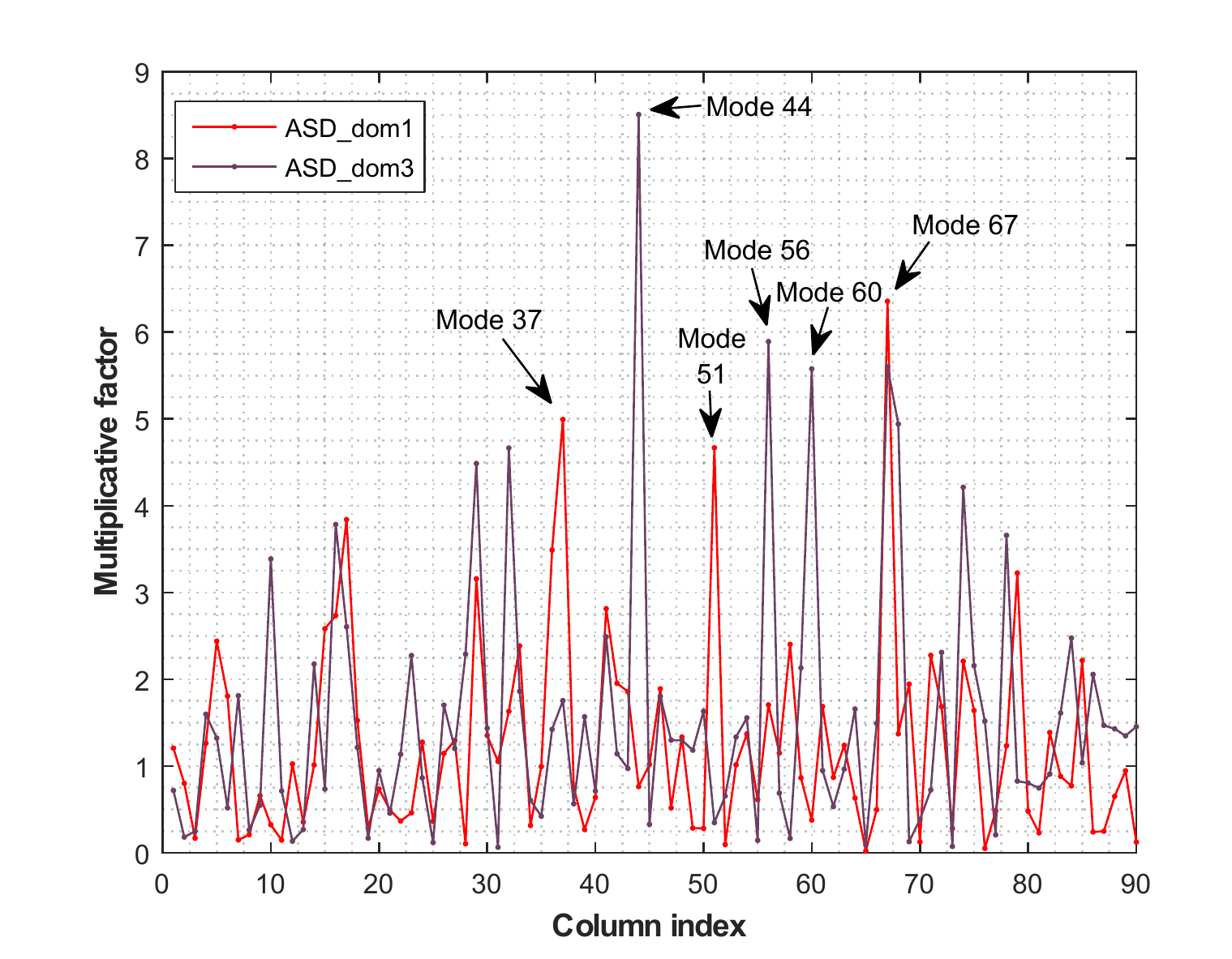}
\caption{Matrix P: interpretation of the rows -- ASD-dominant dimensions.}\label{fig:matP_ASDrows}
\end{figure}

\revision{The results obtained from the proposed framework  confirm that considering the structure-function interplay is crucial in classifying ASD and NT subjects. This pertinent information is however hidden in discriminative patterns which are made of complex combinations of the graph Fourier modes. As explained in Sec.~\ref{sec:lastStep}, the analysis of the projection matrix $\mathbf{P}$ allows us to understand how these combinations are made, and which modes contribute the most to the discriminative features. Figs.~\ref{fig:matP_NTrows} and~\ref{fig:matP_ASDrows} show the values of the matrix $\mathbf{P}$ rows (in absolute values) which correspond to the dimensions considered by the decision tree of Fig.~\ref{fig:DecisionTree}.}

Fig.~\ref{fig:matP_NTrows} is related to the set of weights applied on the GFT coefficients for their projection on the NT-dominant dimensions used by the predictive model (see Fig.~\ref{fig:DecisionTree}). We notice that some weights are dominant over the others of the same raw, i.e., some GFT coefficients are, in terms of absolute values, more amplified than others in the final projection space. This allows to point out the corresponding graph Fourier modes, belonging to different levels of frequency (i.e., low, medium, high), as key patterns for classification, whose corresponding weights are deviating from the mean by more than 2.5 times the standard deviation. The same observation is valid in Fig.~\ref{fig:matP_ASDrows} for the ASD-dominant dimensions. 

Given the above, we are lead to an interesting interpretation of the results. Indeed, by computing the variance of the projected GFT coefficients, we actually measure the variability over time of the presence of some graph Fourier modes in the fMRI signals. These modes may be seen as \textit{frequency signatures} of the NT/ASD conditions. This constitutes another difference with the SFM method which allows to point out isolated prominent regions~\citep{Subbaraju2017}. 

\begin{figure*}
\centering
\begin{tabular}{cc}
\includegraphics[scale=0.25]{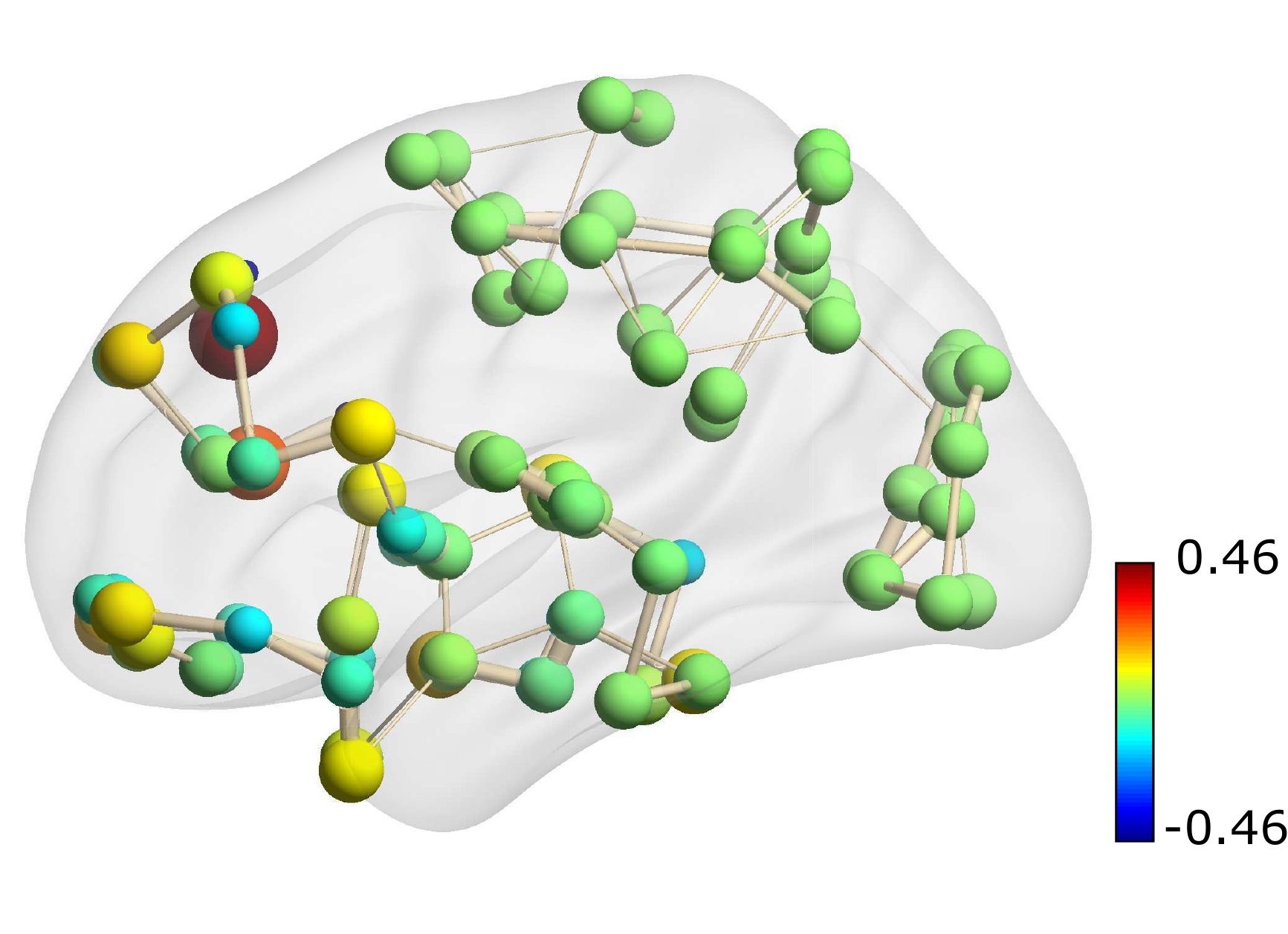} ($M_{49}$) & \includegraphics[scale=0.25]{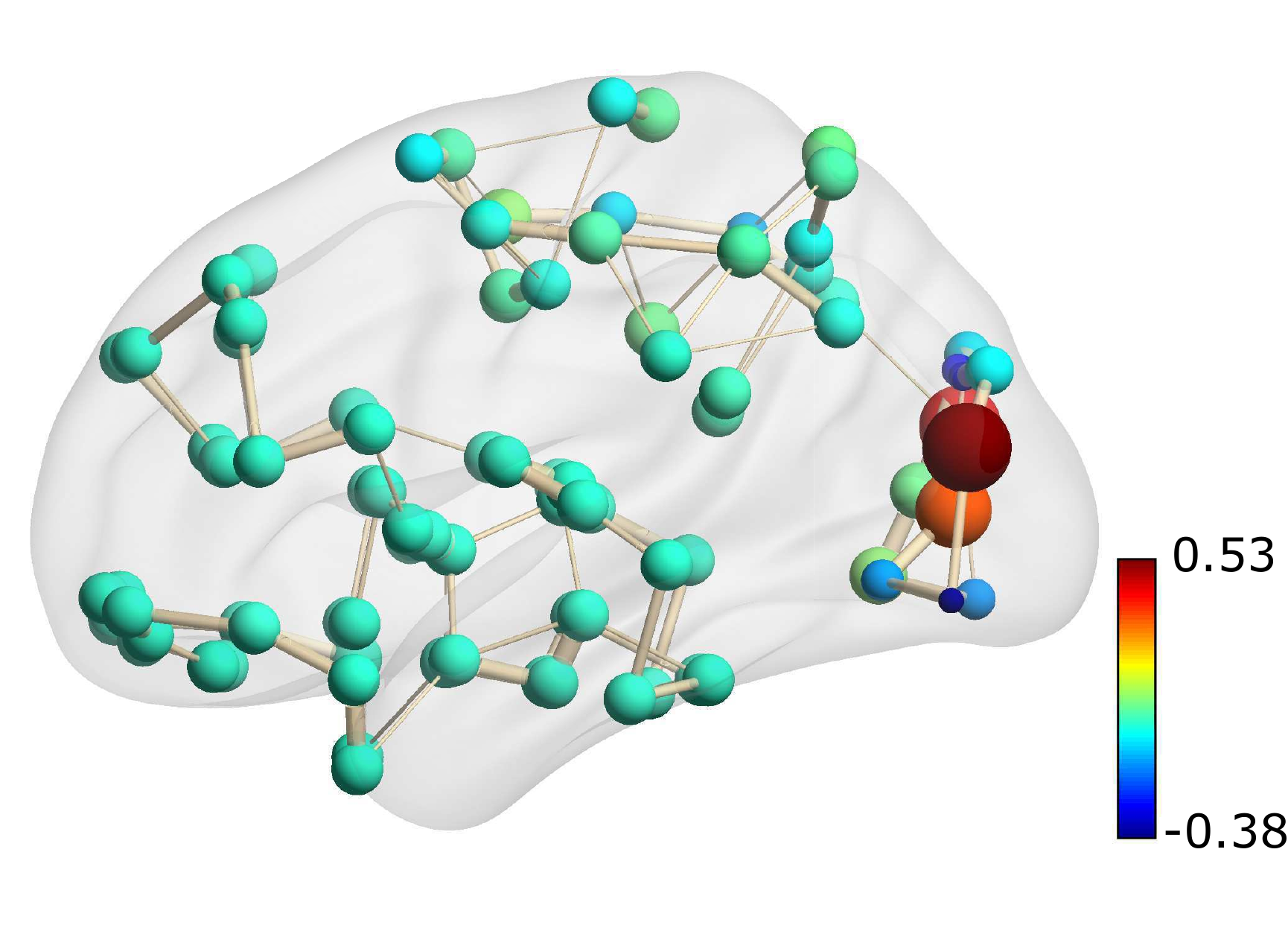} ($M_{50}$) \\ 
\includegraphics[scale=0.25]{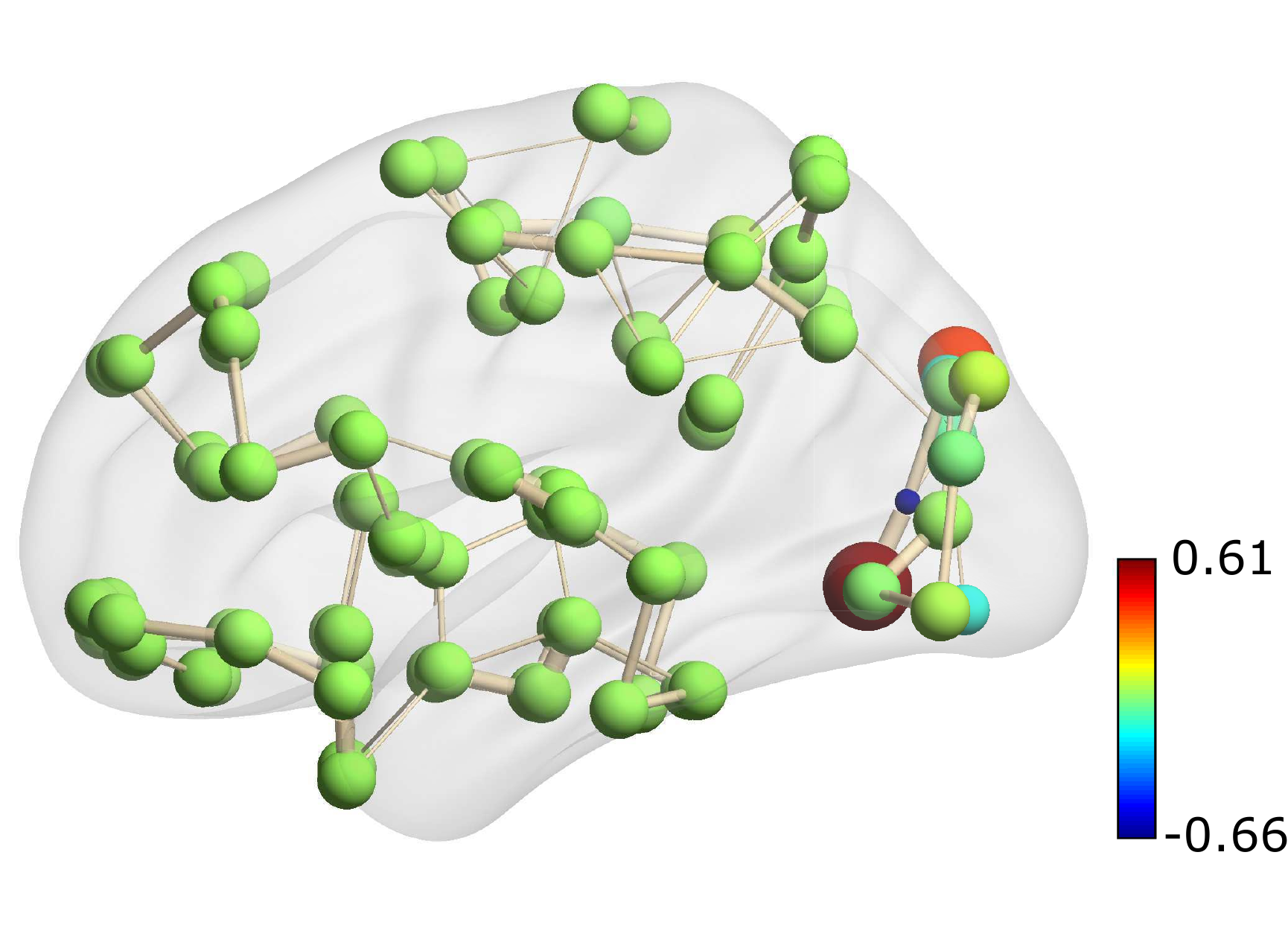} ($M_{70}$) & \includegraphics[scale=0.25]{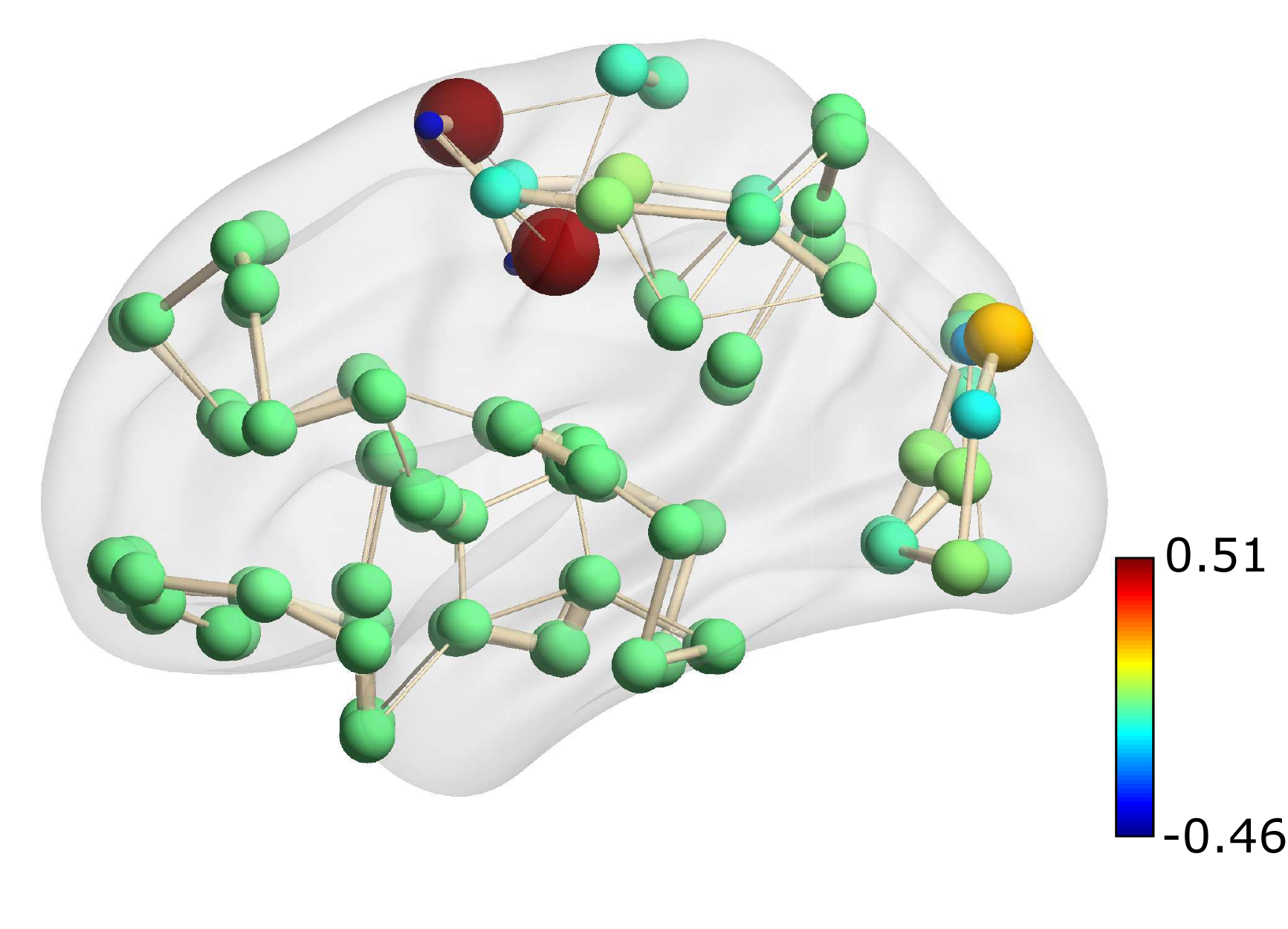} ($M_{79}$) \\ 
\end{tabular}
\caption{Significant frequency Modes ($M$) in NT patients.}\label{fig:freqModesNT}
\end{figure*}

\begin{figure*}
\centering
\begin{tabular}{cc}
    \includegraphics[scale=0.25]{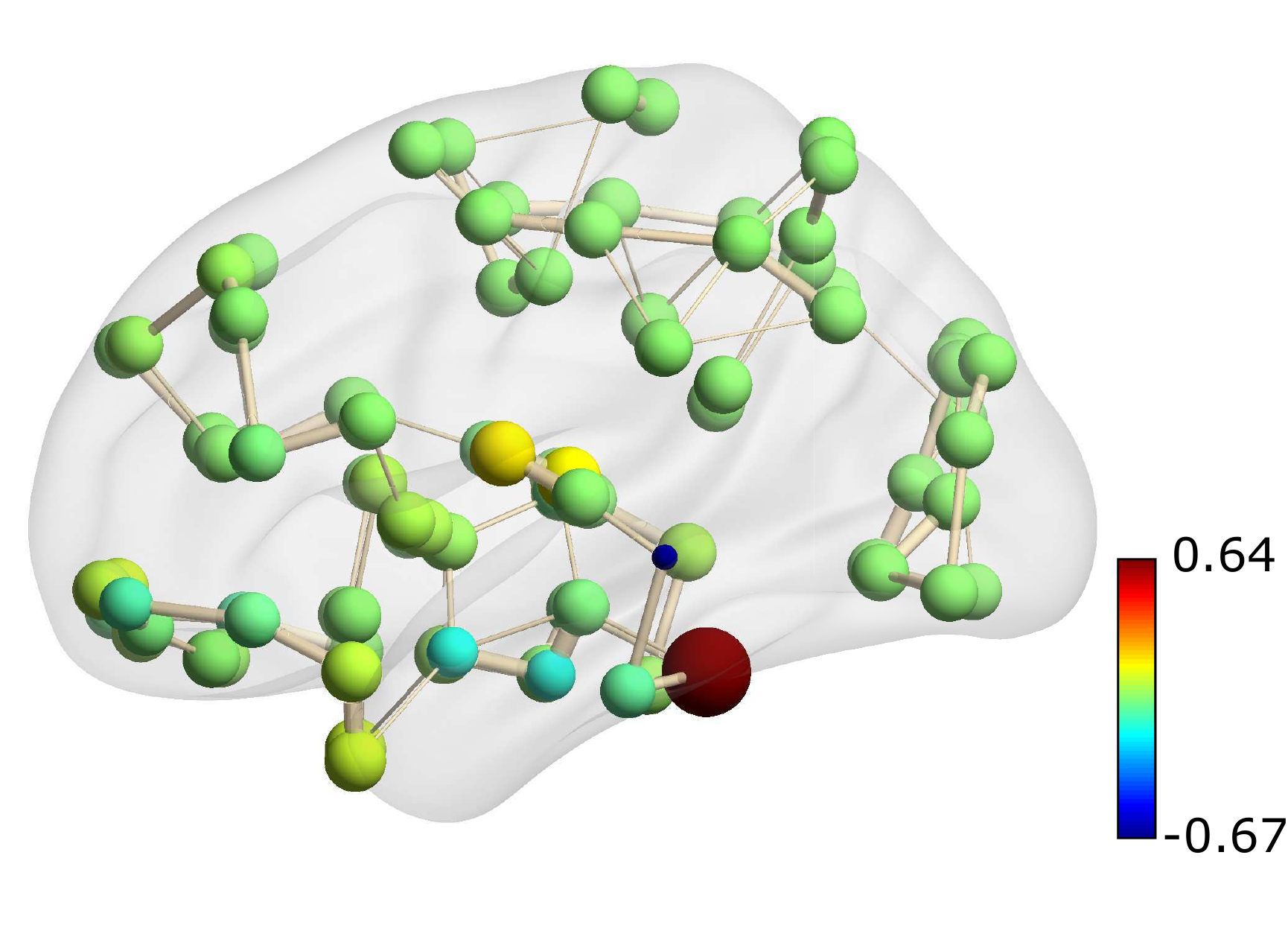} ($M_{37}$) & \includegraphics[scale=0.25]{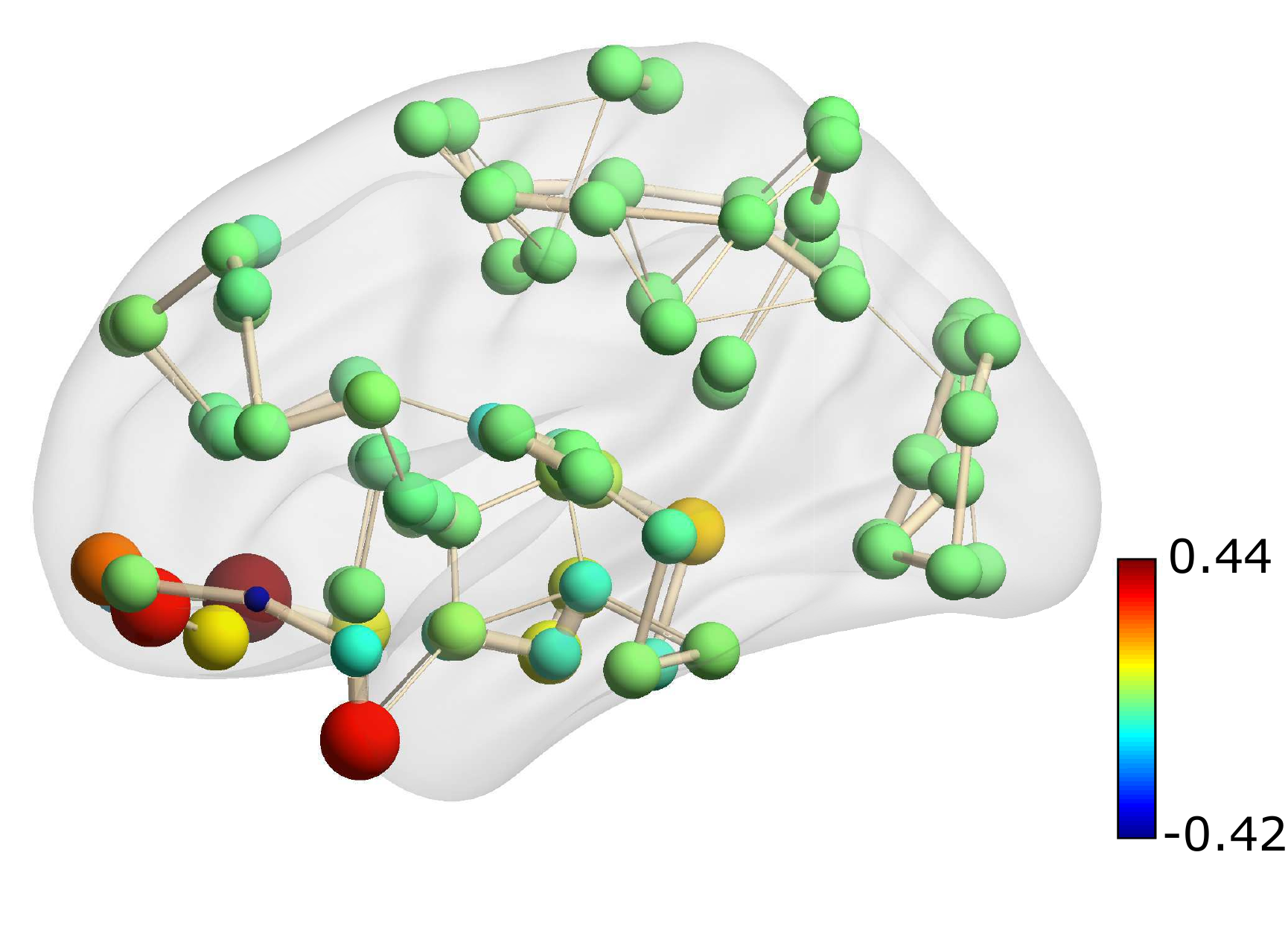} ($M_{44}$) \\ \includegraphics[scale=0.25]{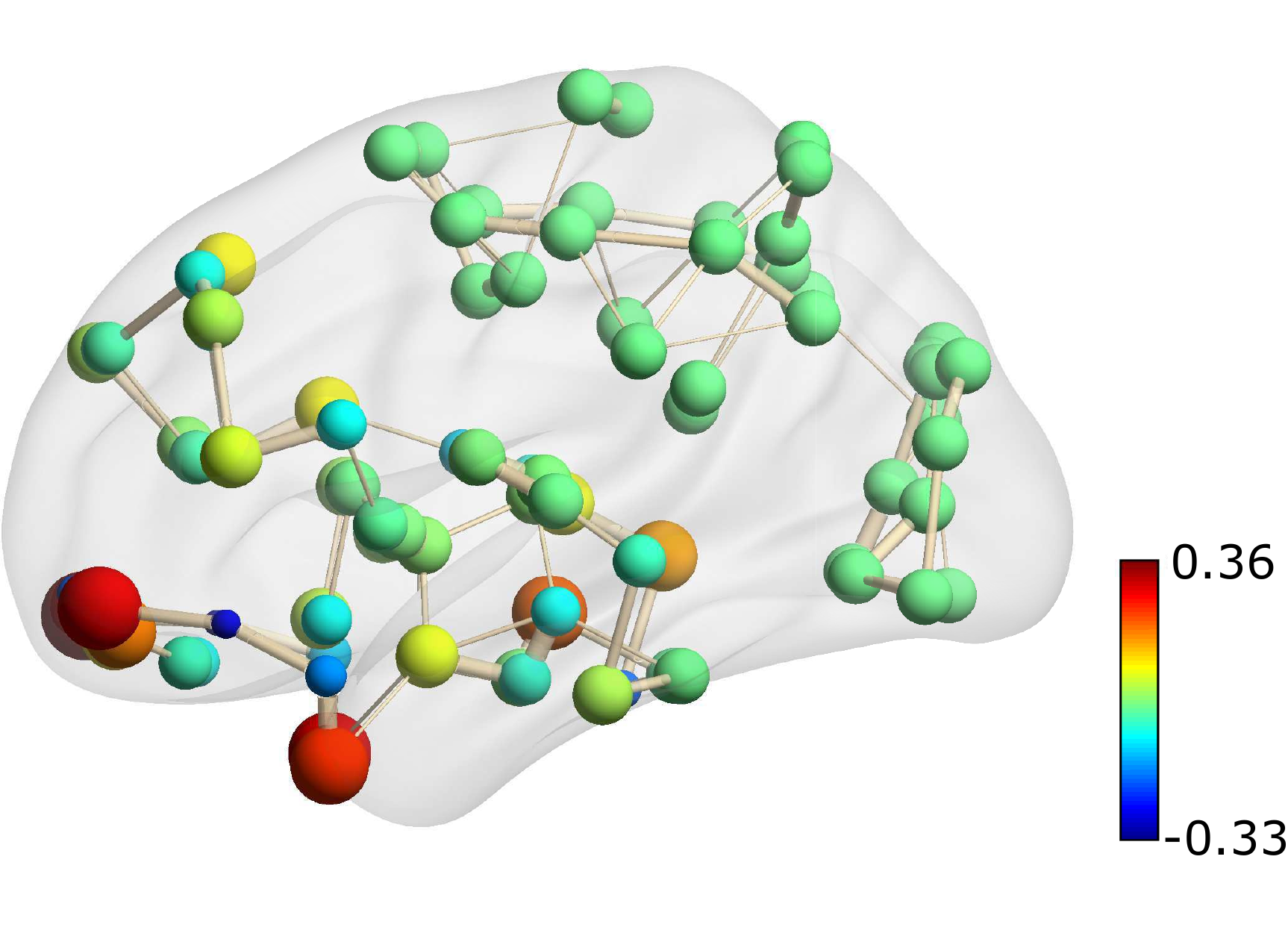} ($M_{51}$) &
    \includegraphics[scale=0.25]{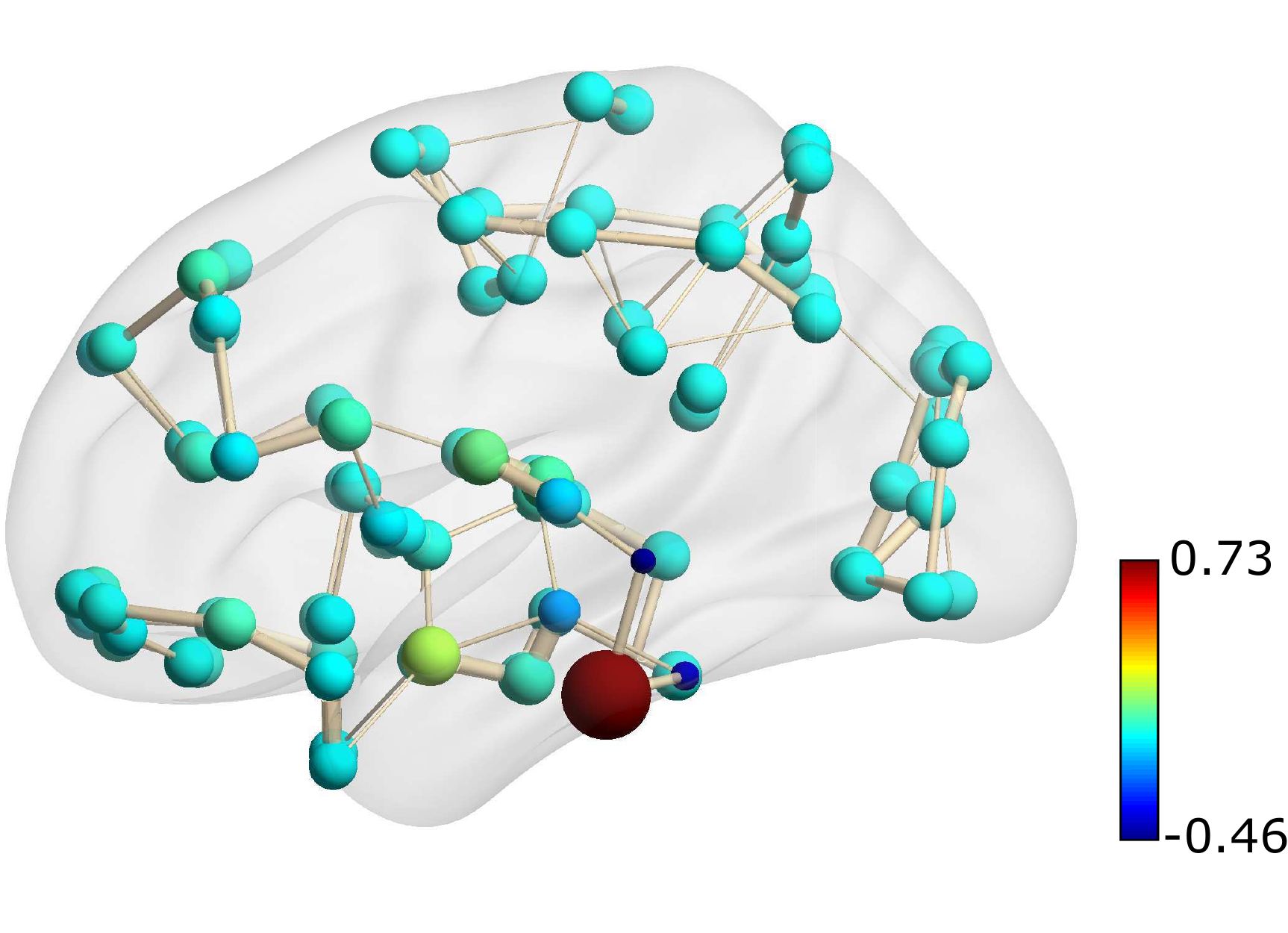} ($M_{56}$)\\ \includegraphics[scale=0.25]{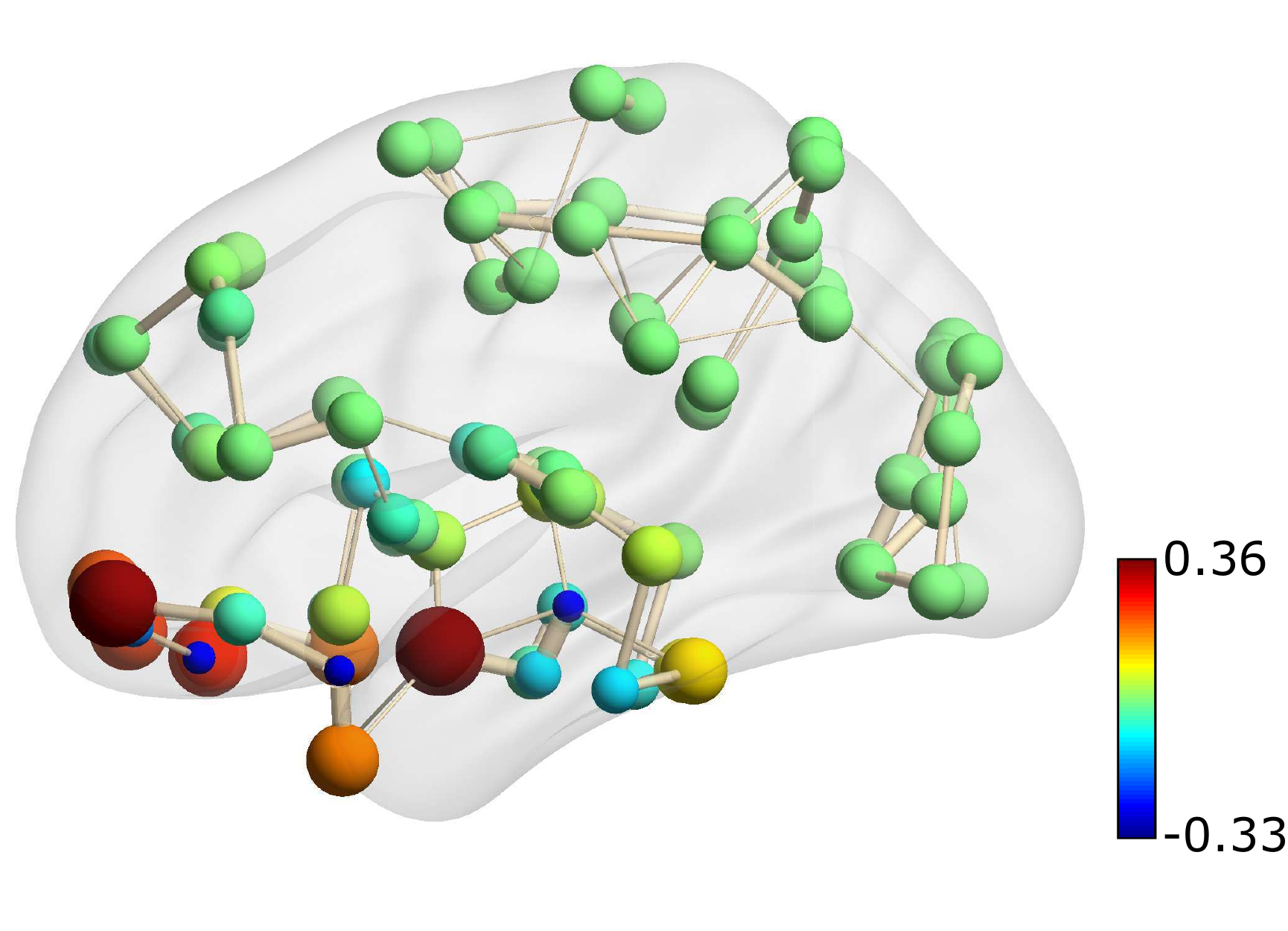} ($M_{60}$)  &\includegraphics[scale=0.25]{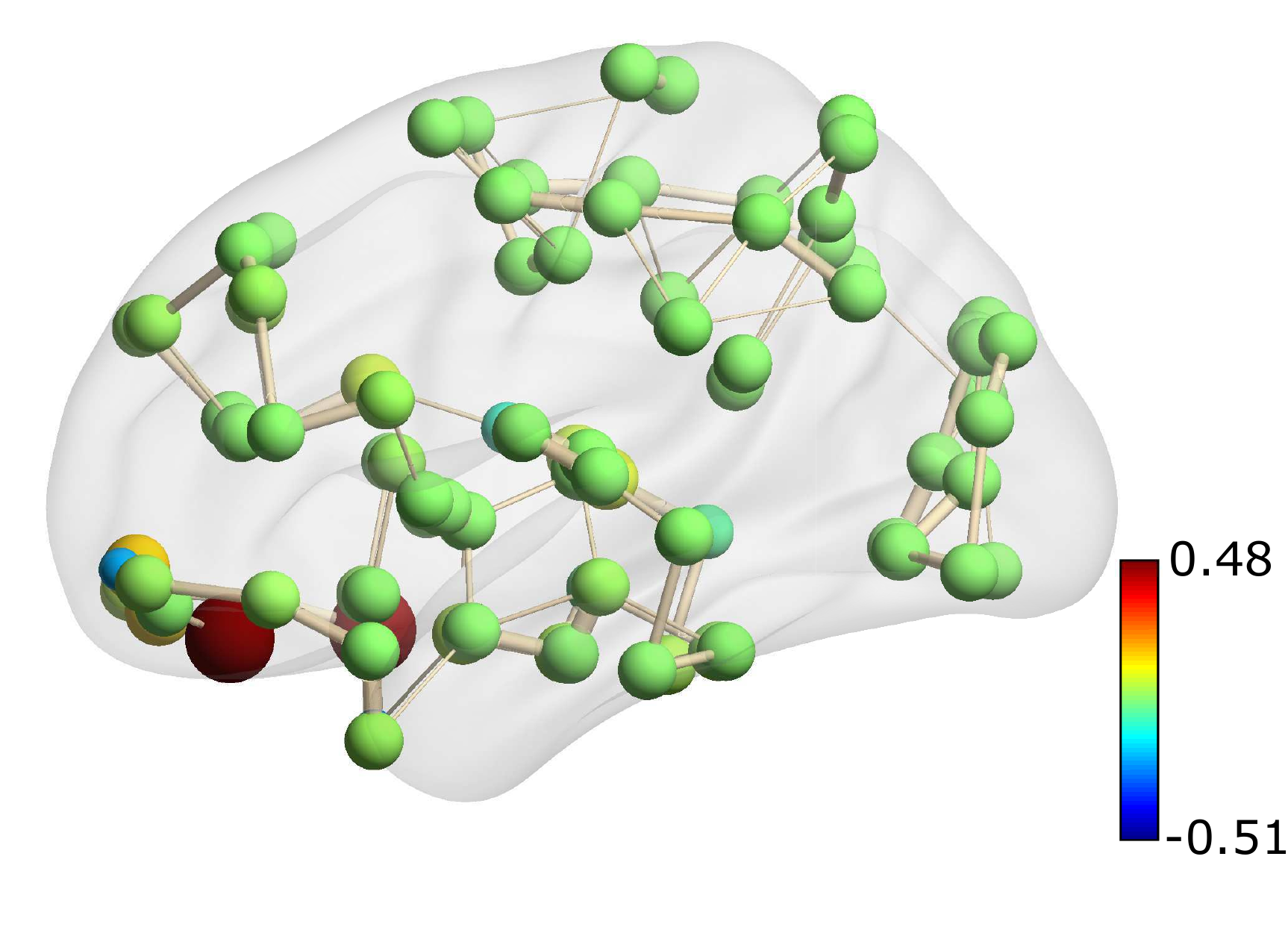} ($M_{67}$)
\end{tabular}
\caption{Significant frequency Modes ($M$) in ASD.}\label{fig:freqModesASD}
\end{figure*}

The analysis of the graph Fourier modes pointed out in Fig.~\ref{fig:matP_NTrows} reveals three out of the four signatures which correspond to a predominant activity in the parieto-occipital regions (see Fig.~\ref{fig:freqModesNT}). Regarding the significant modes of the ASD population which are pointed out in Fig.~\ref{fig:matP_ASDrows}, they are all related to high levels of activity in the fronto-temporal areas in ASD subject (see Fig.~\ref{fig:freqModesASD}). This result is consistent with the previous findings reported in the literature of neuroscience about the influence of the frontal and temporal lobes in ASD~\citep{Hirata2018, Lauvin2012, Poustka2012, Goldberg1999}. 

\subsection{\revision{Comparison with the literature}}~\label{sec:compABIDE}
\revision{The availability of the ABIDE dataset has surely made a major contribution in the increasing literature about data-driven ASD prediction, based on rs-fMRI data. In this respect, the last few years have seen a significant interest for deep learning algorithms in attempt to predict ASD  more accurately~\citep{Lu2020, Eslami2019, Gazzar2019, Heinsfeld2018}. Recent works though  tend to show that it may be possible to reach similar performances through less complex classification schemes such as linear SVMs~\citep{Thomas2020}. The SVM classifier has also been commonly considered in several previous studies~\citep{Kazeminejad2019, Subbaraju2017}. The linear version of the classifier is notably appreciated for its interpretability~\citep{Abraham2017}}.

\revision{The progress achieved on the ABIDE dataset is difficult to estimate from the available literature since the works differ in many respects, e.g.~definition of inclusion criteria, segmentation in training and test sets, feature extraction/selection~\citep{Itani2019b}. However, we can compare the present work with the literature from a general perspective and raise the following elements.}
\begin{itemize}
    \item \revision{Our framework yields favorable performances in comparison to the work of~\cite{Abraham2017} which also aimed at achieving a form of interpretability, through the use of a linear SVM. This indicates the strength and pertinence of the approach that we propose for the extraction of discriminative features.}
    \item \revision{The reported performances are getting close to 75\% based on a decision tree, which is definitely promising while deep learning architectures and SVMs are the most common classifiers used in the literature related to the ABIDE dataset. Indeed, decision trees present a simple and readable structure in comparison to deep learning models. In comparison to SVMs, decision trees are directly focused on the most relevant training features, and provide explanations in the form of logical sequences, thus raising the interactions between the features.}
    \item \revision{The results show how the structural information may influence and improve the performance achieved by using only  fMRI data. This interesting finding further develops the approach consisting of using functional connectivity, which has been commonly considered in the study of ASD and the ABIDE dataset so far (see e.g.,~\citealt{Kazeminejad2019, Dammu2019, Heinsfeld2018,Abraham2017, Subbaraju2017}).}
\end{itemize}

\subsection{\revision{Limitations and future directions}}\label{sec:limitations}

\revision{In terms of both performance and interpretation, the present study yields strong results given (i) the heterogeneity of the samples, and (ii) the simplistic definition of the brain topology. Indeed, our sample is heterogeneous by the range of the patients' ages (between 6 and 18 years old), the multisite nature of the data (seventeen sites contributed to the ABIDE-I dataset) and the range of profiles covered by the ASD spectrum. This disparity may incidentally explain the variability existing around the reported accuracies~\citep{Abraham2017}.  Studying larger and more homogeneous samples merits further investigations.}

\revision{Moreover, the topological structure that we consider is only an approximation of the brain structural connectivity. The assessment of our approach against true topologies, i.e., deduced from individual DTI-based matrices, possibly thresholded to keep the most significant connections, is definitely interesting. The discriminative features would still be analyzed to raise influential frequency modes.  They would however be unique to each person in terms of structural connectivity.}

\section{Conclusion}\label{sec:conclusion}

\revision{In this work, we extended the Graph Signal Processing (GSP) framework by introducing a new algorithm which classifies time-varying graph signals, with application to the identification of Autism Spectrum Disorder (ASD). The algorithm exploits the structure (i.e., brain topology) - function (i.e., fMRI-based activity) interplay of the brain to predict Autism Spectrum Disorder (ASD). In this context, we defined a brain graph as a set of regions of interest connected in terms of their topological distance. For such a purpose, we used the Graph Fourier Transform (GFT) which generalizes the notion of frequency analysis in an irregular domain. The resulting features were processed through an extension of the Fukunaga-Koontz transform to build discriminative markers for the classification of ASD and neurotypical patients. The approach was applied on the publicly available ABIDE dataset.} We summarize below the main conclusions of this study. 
\begin{itemize}
    \item  The proposed methodology performed favorably in comparison to state-of-the-art methods, on the basis of a decision tree as a predictive model. 
    \item The analysis of the results reveals the influence of the frontal and temporal lobes in the diagnosis of the disorder. This finding is consistent with previous reports of the literature of neuroscience. 
    \item In terms of brain activity, we show that the neuropathology may not be attributed to impairments in \revision{only} low, medium and high frequency modes. Indeed, we observe the significant involvement of components that are picked in different parts of the frequency spectrum in the creation of the discriminative features.    
\end{itemize}
These findings indicate that exploiting jointly structural and functional information  of  the brain regions is clearly \revision{a direction to be pursued in the study of ASD}. We believe that our approach paves the way for a better understanding of the disease and thus the exploration of new research avenues by medical experts. 

\section*{\revision{Conflicts of interest}} 
\revision{None.}

\section*{Acknowledgements}
Sarah Itani is a research fellow of the F.R.S.-FNRS. We thank Professors Philippe Fortemps, Arnaud Vandaele, Fabian Lecron (Faculty of Engineering, University of Mons), Professor Mandy Rossignol (Faculty of Psychology and Education, University of Mons, Belgium), Dr. Xiaowen Dong (University of Oxford), and Dr. Christine Choirat (Swiss Data Science Center - EPFL and ETH Zurich) for their advice and interest in this work. 
\appendix
\section{Appendices}

\subsection{The pair-wise joint expectancy is semi-definite positive}\label{appA}

We can prove that the pair-wise joint expectancy $\mathbf{S}_i$, as defined by Eq.~\ref{eq:jointExp}, is positive semi-definite. For simplicity, we ignore the index $i$. By definition, a matrix $\mathbf{S}$ is positive semi-definite if: 
\begin{align*}
x^{T} \mathbf{S}\: x \geq 0 \:\:\:\: \forall x \in \mathbb{R}^{n}. 
\end{align*}
In this case, $\mathbf{S} = \mathbf{Y}\mathbf{Y}^{T}$, which involves, $\forall x \in \mathbb{R}^{n}$: 
\begin{align*}
x^{T} \mathbf{S}\: x = x^{T} \mathbf{Y}\mathbf{Y}^{T}\: x = (\mathbf{Y}^{T}\: x)^{T} (\mathbf{Y}^{T}\: x) = \| \mathbf{Y}^{T}\: x \| ^2 \geq 0. 
\end{align*}
\subsection{The mean pair-wise joint expectancy is non invertible}\label{appB}
Let us express the matrix $\mathbf{Y}_i$ of normalized GFT coefficients for a given patient $i$ as: 
\begin{equation}\label{Normalization}
    \mathbf{Y}_i = (\mathbf{\hat{X}}_i - \mathbf{M}_i)\cdot \mathbf{R}_i \:\:\:\: \text{with}\:\:\:\: \mathbf{M}_i = \mathbf{U} \mathbf{V}^{T}\mathbf{X}_i 
\end{equation}
where $\mathbf{R}_i$ is a diagonal matrix including the norm of the columns of $\mathbf{Y}_i$ and $\mathbf{M}_i$ is a matrix including the mean of the columns of $\mathbf{\hat{X}}_i$; $\mathbf{U}$ is a $r\times r$ constant matrix including entries equal to $1/r$.

Given Eq. \ref{Normalization}, an alternative expression for the mean joint expectancy matrix $\mathbf{\Bar{S}}$ is:  
\begin{align*}
    \mathbf{\Bar{S}} & = 
    \frac{1}{n}\sum_i (\mathbf{\hat{X}}_i - \mathbf{M}_i)\cdot \mathbf{H}_i\cdot (\mathbf{\hat{X}}_i - \mathbf{M}_i)^{T}.
\end{align*}
where, for the sake of simplicity, we denote $\mathbf{H}_i$ as:
\begin{equation*}
    \frac{{\mathbf{D}_i}^{2}}{\mathrm{Tr}(\mathbf{Y}_i{\mathbf{Y}_i}^{T})}.
\end{equation*}

The matrix $\mathbf{\Bar{S}}$ has a zero eigenvalue associated to a constant eigenvector, that is, if $q$ denotes a constant column vector:  
\begin{equation*}
\mathbf{\Bar{S}} q = 0. 
\end{equation*}
***
\textit{Proof}
\begin{align}\label{globJointExp}
\mathbf{\Bar{S}}q&= \frac{1}{n}\sum_i (\mathbf{\hat{X}}_i  - \mathbf{M}_i)\cdot \mathbf{H}_i\cdot(\mathbf{\hat{X}}_i^{T} - \mathbf{M}_i^{T}) \cdot q
\end{align}
Let us focus on the individual terms, i.e., for each $i$ : 

$(\mathbf{\hat{X}_i} - \mathbf{M}_i)\cdot \mathbf{H}_i \cdot(\mathbf{\hat{X}}_i^{T} - \mathbf{M}_i^{T})\cdot q$
\begin{align}\label{ScatMat}
=  \mathbf{\hat{X}}_i \mathbf{H}_i \mathbf{\hat{X}}_i^{T} q - \mathbf{\hat{X}}_i \mathbf{H}_i \mathbf{M}_i^{T} q - \mathbf{M}_i \mathbf{H}_i \mathbf{\hat{X}}_i^{T} q + \mathbf{M}_i \mathbf{H}_i \mathbf{M}_i^{T} q. 
\end{align}
Yet $\mathbf{\hat{X}}_i^{T} q = \mathbf{M}_i^{T} q$. Indeed:
\begin{align*}
\mathbf{M}_i^{T} q &= (\mathbf{U} \mathbf{V}^{T}\cdot \mathbf{X}_i)^{T} q \\
&= (\mathbf{V}^{T}\cdot \mathbf{X}_i)^{T} \mathbf{U}^{T} q \\
&=  \mathbf{\hat{X}}_i^{T} \mathbf{U}^{T} q.
\end{align*}
As $\mathbf{U}$ is a square constant matrix, $\mathbf{U}^{T} = \mathbf{U}$. The vector $\mathbf{U} q$ includes the mean of each element of the vector $q$, which is a constant vector. Thus $\mathbf{U}q = q$ and
\begin{equation*}
\mathbf{M}_i^{T} q =  \mathbf{\hat{X}}_i^{T} q
\end{equation*} 
which means that \eqref{ScatMat} sums to zero and thus, \eqref{globJointExp} also. This is related to the mean centering operation, executed over the columns of the matrices $\mathbf{\hat{X}}_i$. 
***
\subsection{The Newcomb's theorem}\label{appC}
\noindent
For the sake of completeness, we recall here the main results of the theorem.
\newtheorem*{mydef}{Diagonalization~\citep{Newcomb1961}}
\begin{mydef}
Let $\mathbf{A}$ and $\mathbf{B}$ be $n\times n$ real, symmetric, positive semi-definite matrices. Then there exists a real non-singular matrix $\mathbf{T}$ and real diagonal matrices $\mathbf{A}_0$ and $\mathbf{B}_0$ such that 
\begin{align*}
\mathbf{A} &= \mathbf{T} \mathbf{A}_0 \mathbf{T}^{T}  \\
\mathbf{B} &= \mathbf{T} \mathbf{B}_0 \mathbf{T}^{T}
\end{align*}
where 
\begin{align*}
\mathbf{A}_0 &= \mathbf{\text{diagonal}[0_{n-a}, I_a]}
\end{align*}
if $\mathbf{a}$ denotes the rank of matrix $\mathbf{A}$. 
\end{mydef}

\subsection{\revision{Simultaneous diagonalization of the whitened matrices}}\label{psdSA} 
\paragraph{\revision{Preliminaries}} \revision{Let us show that $\mathbf{\bar{S}^{A'}}$ and $\mathbf{\bar{S}^{N'}}$ are positive semi-definite (\textit{psd}) and have the following structures: 
\[ \mathbf{\bar{S}^{A'}} = 
\setlength\arraycolsep{5pt}
\left[\begin{array}{c|c}
\multicolumn{2}{c}{0 \:\:\: \ldots \:\: 0} \\ 
\cline{2-2}
\vdots & \multirow{2}{*}{$\mathbf{\bar{S}^{A'}_{r-1}}$}\\
0 & \\
\end{array}\right] \:\:\:\:\:\:\:\:\:\:
 \mathbf{\bar{S}^{N'}} = 
\setlength\arraycolsep{5pt}
\left[\begin{array}{c|c}
\multicolumn{2}{c}{0 \:\:\: \ldots \:\: 0} \\ 
\cline{2-2}
\vdots & \multirow{2}{*}{$\mathbf{\bar{S}^{N'}_{r-1}}$}\\
0 & \\
\end{array}\right].
\]
We present here the reasoning for $\mathbf{\bar{S}^{A'}}$. The same applies to $\mathbf{\bar{S}^{N'}}$. }

\revision{By definition, as the matrix of joint expectancy $\mathbf{\bar{S}^{A}}$ is \textit{psd}, we have: 
\begin{align*}
x^{T} \mathbf{\bar{S}^{A}} x \geq 0 \:\:\:\: \forall x \in \mathbb{R}^{n}. 
\end{align*}
We can show that $\mathbf{\bar{S}^{A'}}$ is \textit{psd} as well. Indeed, let us set $x = \mathbf{Q}_2^{T}y$, $\forall y \in \mathbb{R}^{n}$. Then:
\begin{align*}
y^{T} \mathbf{\bar{S}^{A'}} y = y^{T} \mathbf{Q}_2 \mathbf{\bar{S}}^A \mathbf{Q}_2^{T} y &= (\mathbf{Q}_2^{T} y)^{T} \mathbf{\bar{S}^A} (\mathbf{Q}_2^{T} y) \\ 
& = x^{T} \mathbf{\bar{S}^A} x \geq 0 
\end{align*}
which involves $y^{T} \mathbf{\bar{S}^{A'}} y \geq 0 \Rightarrow \mathbf{\bar{S}^{A'}} \succeq 0$. As a \textit{psd} matrix, the diagonal entries of $\mathbf{\bar{S}^{A'}}$ are positive~\citep{Golub2012}. 
Thus, to satisfy Eq. \ref{diag}, the first main diagonal entry of $\mathbf{\bar{S}^{A'}}$ must be equal to zero; the corresponding row and column are zero, given the \textit{psd-ness} of $\mathbf{\bar{S}^{A'}}$ (resp. $\mathbf{\bar{S}^{N'}}$)~\citep{Horn1990, Golub2012}. }

\paragraph{\revision{Diagonalization}} \revision{In order to diagonalize $\mathbf{\bar{S}^{A'}}$,~\cite{Newcomb1961} proposes to diagonalize  $\mathbf{\bar{S}^{A'}_{r-1}}$ by an orthogonal transformation $\mathbf{T'}$ that we deduce through eigen-decomposition. The global transformation matrix $\mathbf{T_2}$  constitutes of the  following: }
\revision{
\[ \mathbf{T_2} = 
\setlength\arraycolsep{5pt}
\left[\begin{array}{c|c}
1 & 0\:\:\: \ldots \:\: 0 \\ \hline 
0 & \multirow{3}{*}{$\mathbf{T^{'}}$}\\
\vdots & \\
0 & \\
\end{array}\right]. 
\]}

\revision{Thus, $\mathbf{T_2^{T} \bar{S}^{A'} T_2} = \mathbf{\bar{S}^{A^{''}}}$, where $\mathbf{\bar{S}^{A^{''}}}$ is a diagonal matrix. Eq.~\ref{diag} can be reformulated as:
\begin{align*}
 \mathbf{T_2^{T}~\text{diagonal}[0, I_{r-1}]~T_2} &=   \mathrm{\alpha_A}  \mathbf{T_2^{T}  \bar{S}^{A'} T_2} + ~\mathrm{\alpha_N}  \mathbf{T_2^{T} \bar{S}^{N'} T_2}  
 \\ 
\Leftrightarrow \: \mathbf{\text{diagonal}[0, I_{r-1}]} &= \mathrm{\alpha_A} \mathbf{\bar{S}^{A^{''}}} + \mathrm{\alpha_N} \mathbf{\bar{S}^{N^{''}}} 
\end{align*}
Given that $\mathbf{\text{diagonal}[0, I_{r-1}]}$ and $\mathbf{\bar{S}^{A^{''}}}$ are diagonal matrices, $\mathbf{\bar{S}^{N^{''}}}$ is diagonalizable, and it shares the same eigenvectors with  $\mathbf{\bar{S}^{A^{''}}}$.}

\subsection{Brain figures}\label{figInfo}

Table~\ref{tab:labelsAAL} lists the ROIs of the AAL atlas grouped by partition, as suggested by~\cite{Wang2012}. These ROIs are labeled with the notations proposed by the \textsc{BrainNet Software}, which was used to draw the brain figures of the present paper. The ROI indexes related to the AAL atlas are also provided: odd (resp. even) numbers refer to ROIs from the left (resp. right) hemisphere. 

\begin{table}
    \centering
    \resizebox{0.5\textwidth}{!}{
    \begin{tabular}{l|l|l|l}
    \toprule 
    \textsc{Partition} & \textsc{Index L-R} & \textsc{ROI}  & \textsc{Label}   \\
    \midrule
        \multirow{10}{*}{Frontal} & 03-04&$\text{Frontal}\_\text{Sup}$&SFGdor \\
        &05-06&$\text{Frontal}\_{\text{Sup}\_\text{Orb}}$&ORBsup \\
        &07-08&$\text{Frontal}\_\text{Mid}$&MFG \\
        &09-10&$\text{Frontal}\_{\text{Mid}\_\text{Orb}}$&ORBmid \\
        &11-12&$\text{Frontal}\_{\text{Inf}\_\text{Oper}}$&IFGoperc \\
        &13-14&$\text{Frontal}\_{\text{Inf}\_\text{Tri}}$&IFGtriang \\
        &15-16&$\text{Frontal}\_{\text{Inf}\_\text{Orb}}$&ORBinf \\
        &23-24&$\text{Frontal}\_{\text{Sup}\_\text{Medial}}$&SFGmed \\
        &25-26&$\text{Frontal}\_{\text{Med}\_\text{Orb}}$&ORBsupmed \\
        &27-28&Rectus&REC \\
        \midrule 
        \multirow{9}{*}{Parietal} &01-02&Precentral&PreCG \\
        &19-20&$\text{Supp}\_{\text{Motor}\_\text{Area}}$&SMA \\
        &57-58&Postcentral&PoCG \\
        &59-60&$\text{Parietal}\_\text{Sup}$&SPG \\
        &61-62&$\text{Parietal}\_\text{Inf}$&IPL \\
        &63-64&SupraMarginal&SMG \\
        &65-66&Angular&ANG \\
        &67-68&Precuneus&PCUN \\
        &69-70&$\text{Paracentral}\_\text{Lobule}$&PCL \\
        \midrule
        \multirow{7}{*}{Occipital} &43-44&Calcarine&CAL \\
        &45-46&Cuneus&CUN \\
        &47-48&Lingual&LING \\
        &49-50&$\text{Occipital}\_\text{Sup}$&SOG \\
        &51-52&$\text{Occipital}\_\text{Mid}$&MOG \\
        &53-54&$\text{Occipital}\_\text{Inf}$&IOG \\
        &55-56&Fusiform&FFG \\
        \midrule
        \multirow{11}{*}{Temporal} &17-18&$\text{Rolandic}\_\text{Oper}$r&ROL \\
        &29-30&Insula&INS \\
        &37-38&Hippocampus&HIP \\
        &39-40&ParaHippocampal&PHG \\
        &41-42&Amygdala&AMYG \\
        &79-80&Heschl&HES \\
        &81-82&$\text{Temporal}\_\text{Sup}$&STG \\
        &83-84&$\text{Temporal}\_{\text{Pole}\_\text{Sup}}$&TPOsup \\
        &85-86&$\text{Temporal}\_\text{Mid}$&MTG \\
        &87-88&$\text{Temporal}\_{\text{Pole}\_\text{Mid}}$&TPOmid \\
        &89-90&$\text{Temporal}\_\text{Inf}$&ITG \\
        \midrule
        \multirow{3}{*}{Cingulum} &31-32&$\text{Cingulum}\_\text{Ant}$&ACG \\
        &33-34&$\text{Cingulum}\_\text{Mid}$&DCG \\
        &35-36&$\text{Cingulum}\_\text{Post}$&PCG \\
        \midrule
        \multirow{5}{*}{Subcortical} &21-22&Olfactory&OLF \\
        &71-72&Caudate&CAU \\
        &73-74&Putamen&PUT \\
        &75-76&Pallidum&PAL \\
        &77-78&Thalamus&THA \\
        \bottomrule 
    \end{tabular}}
    \caption{AAL atlas~\citep{Xia2013,Wang2012,Tzourio2002}}
    \label{tab:labelsAAL}
\end{table}

\end{document}